\documentclass[12pt]{iopart}
\usepackage{iopams}
\usepackage{epsfig}

\begin{document}

\tolerance = 10000


\title{Ground-state two-spinon bonds in the Hubbard model on a square lattice}
\author{J. M. P. Carmelo} 
\address{GCEP-Center of Physics, U. Minho, Campus Gualtar,
P-4710-057 Braga, Portugal\\
E-mail: carmelo@fisica.uminho.pt}

\begin{abstract}
In this paper the spin configurations of the ground state and one- and two-electron 
excited states of the Hubbard model on the square lattice are studied. 
We profit from a general rotated-electron description, which is
consistent with the model global $SO(3)\times SO(3)\times U(1)$ symmetry.
For rotated electrons, doubly and single occupancy are good quantum
numbers for on-site repulsion $U>0$. The above states are within
that description generated by occupancy configurations of charge $c$ fermions
and spin-singlet two-spinon $s1$ bond particles. Those describe
the charge and spin degrees of freedom, respectively, of the rotated electrons
that singly occupy sites. While the $c$ fermions have no internal structure, that 
of the spin-neutral $s1$ bond-particle occupancy configurations is here 
described in terms of spinon occupancies of a well-defined effective spin lattice. 
In reference \cite{companion} it is confirmed that our results contribute to the further
understanding of the role of electronic correlations in the spin spectrum
of the parent compound La$_2$CuO$_4$. They are also of interest for studies 
of ultra-cold fermionic atoms on an optical lattice.
 \end{abstract}
\pacs{71.10.Fd, 75.10.Jm, 75.10.Lp, 71.10.Pm}

\maketitle

\section{Introduction}

The Hubbard model on a square lattice is the simplest toy model 
for the description of the effects of electronic correlations in the spin
spectrum of the Mott-Hubbard insulators cuprates parent 
compounds \cite{LCO-neutr-scatt,companion}. On the square and cubic lattices the model has 
no exact solution and many open questions about its properties remain unsolved. Fortunately, it
can be experimentally realized with unprecedented precision in systems of correlated ultra-cold fermionic 
atoms on an optical lattice \cite{Zoller}. Recent experiments considered the Hubbard
model on the cubic lattice \cite{cubic}. They involved ultra-cold fermionic atoms on an 
optical cubic lattice formed by interfering laser fields. Similar experiments referring
to the model on the square lattice are in progress. One may expect very detailed 
experimental results over a wide range of parameters to be available.

In this paper we study the spin occupancy configurations of the ground state
of the Hubbard model on a square lattice with $N_a^2\gg 1$ sites within the 
spin-singlet two-spinon $s1$ bond-particle description also 
used in the investigations of Ref. \cite{companion}. 
Here $N_a$ is the number of sites of an edge of length $L=N_a\,a$
and spacing $a$. Hence the total number of sites of the square lattice
of area $L^2$ is $N_a^2\equiv N_a\times N_a$. That description involves the 
generalization to all finite values of the on-site
repulsion $U>0$ of the exact transformation for separation of spin-$1/2$ 
fermions without constraints introduced for large $U$ values
in Ref. \cite{Ostlund-06}. It
is consistent with the global $SO(3)\times SO(3)\times U(1)=[SO(4)\times U(1)]/Z_2$ symmetry
found recently in Ref. \cite{bipartite} for the model
on any bipartite lattice. That global symmetry is an extension 
of the $SO(4)$ symmetry known to occur for
the model on such lattices \cite{Zhang}. 
The extended global symmetry is related to the rotated 
electrons that for $U>0$ emerge
from the electrons through a unitary transformation
of the type considered in Ref. \cite{Stein} and to
the local symmetries and unitary transformations
considered in Ref. \cite{U(1)-NL}. For rotated electrons double
and single occupancy are good quantum numbers for $U>0$.  

Here we study the spin configurations of the model ground states and 
excited states that span the subspace relevant for the one- and 
two-electron physics. It is spanned by an initial 
vanishing-spin-density ground state and the energy and momentum eigenstates contained
in the excitations generated by application onto that
ground state of one- and two-electron operators. Within the description used in
the studies of Ref. \cite{companion}, for $N_a^2\gg 1$ such states can be generated by suitable
occupancy configurations of two basic quantum objects:
spin-less and $\eta$-spin-less charge $c$ fermions and
spin-$1/2$ spinons. There are also $\eta$-spin-$1/2$ $\eta$-spinons, yet
in that subspace they refer to a single occupancy configuration and
thus play no active role. The spinons describe the spin degrees
of freedom of the rotated-electron singly occupied sites. Their spin
projection is the same as that of the corresponding rotated electron.
The $\eta$-spinons of $\eta$-spin projection down and up describe the $\eta$-spin degrees
of freedom of the rotated-electron doubly occupied and unoccupied sites,
respectively. In our subspace rotated-electron double occupancy 
vanishes and there are only $\eta$-spin-up $\eta$-spinons. 
The spinon spin-singlet configurations involve in general the $2\nu$-spinon
and $2\nu$-site $s\nu$ bond particles \cite{companion}. 
Here $\nu=1,2,3,...$ refers to the number of spinon pairs in each $s\nu$ bond. 
Fortunately, the ground states and their one- and two-electron excited states 
considered here, except the
spin-singlet states, involve only two-spinon and two-site $s1$ bond particles.
Moreover, the latter states involve in addition to the $s1$ bond particles
a single four-spinon and four-site $s2$ bond particle.

The $c$ fermions live on a $c$ effective lattice identical to the original lattice.
The $c$ fermion occupied sites describe the charge degrees of freedom to 
those singly occupied by rotated electrons. The corresponding spin degrees
of freedom are described by the spinons. In turn, the $c$ fermion unoccupied 
sites correspond to those doubly occupied or unoccupied by rotated electrons. 
The corresponding $\eta$-spin degrees of freedom are described by the $\eta$-spinons. 
Importantly, the spatial coordinates of the $c$ fermion occupied and 
unoccupied sites are also the spatial coordinates in the original lattice of the
sites referring to spinons and $\eta$-spinons, respectively. 
Provided that $N_a^2\gg 1$ this is behind the occupancies of the spin effective
lattice and $\eta$-spin effective lattice being independent
\cite{companion}. Such effective lattices
involve only the rotated-electron singly occupied sites and the rotated-electron
doubly occupied and unoccupied sites, respectively. 

The spinons that are not part of $s\nu$ bond particles are called
independent spinons. They are invariant under the
electron - rotated-electron unitary transformation. For the lowest-weight states
whose spin $S_{s}$ and spin projection $S^z_{s}$ are such that 
$S_{s}=-S^z_{s}$ the independent spinons have up spin projection.
For the subspace considered here all $\eta$-spinons have up $\eta$-spin projection
so that the $\eta$-spinon effective lattice plays no active role. In turn, since the location in the
original lattice of the sites of the spin effective lattice is recorded in the $c$ fermion occupancy 
configurations, it turns out that provided that $N_a^2\gg 1$ it is a good approximation to
consider that the spin effective lattice is a square lattice with the same length edge $L$ as 
the original lattice and a number of sites $N_{a_{s}}^2$ and spacing $a_{s}$ given by,
\begin{equation}
N_{a_{s}}^D = (1-x)\,N_a^D
\, ; \hspace{0.35cm}
a_{s} = {a\over (1-x)^{1/D}} \, ,
\hspace{0.25cm} (1-x)\geq 1/N_a^D \, .
\label{NNCC}
\end{equation}
Here $D=2$ refers to the present square lattice, $D=1$ corresponds
to the one-dimensional (1D) lattice some times considered below, and
$x=(N_a^D -N)/N_a^D$ is the hole concentration.
For $N_a^2\gg 1$, in spite of rotated-electron single occupancy being only conserved
globally and not locally, the spinon occupancies of the spin effective lattice provide a
good approximation of the ground state and excited states spin configurations \cite{companion}. 
Furthermore, within the present description such configurations involve 
the occupancies of a $s1$ effective lattice. Its unoccupied sites
are the up-spin independent spinons. Its occupied sites refer to the two-spinon $s1$ bond particles 
studied in this paper. Hence each occupied site of the $s1$ effective lattice corresponds to two sites 
of both the original lattice and spin effective lattice.

Bond states based on two-spinon and two-site bonds are in general 
not orthogonal and their basis is 
overcomplete. However, within the representation used
in the studies of this paper such a problem does not occur.
This is due to physical restrictions of the occupancy configurations
of the well-defined set of two-site bonds contributing to a 
given $s1$ bond particle. For instance, that 
they are centered at the same real-space 
coordinate of the spin effective lattice. For 1D the discrete momentum
values of both the $c$ fermions and $s1$ fermions generated from
the $s1$ bond particles studied here through an extended
Jordan-Wigner transformation \cite{companion,J-W,Wang,Feng}
are good quantum numbers whose occupancy
configurations generate the energy eigenstates \cite{1D}. 
   
Evidence that for the one- and two-electron subspace the states 
generated by the occupancy configurations of such discrete 
momentum values are for the model on the square lattice energy
eigenstates is given in the related investigations carried out in Ref. 
\cite{companion}. The square-lattice quantum liquid introduced in that
reference contains the one- and two-electron excitations of the Hubbard model on a square lattice.
At hole concentration $x=(N_a^2-N)/N_a^2=0$, $U/4t\approx 1.525$, and $t\approx 295$ meV 
it is found in that reference to quantitatively describing the spin-wave spectrum observed in the parent compound 
La$_2$CuO$_4$ \cite{LCO-neutr-scatt}. In turn, it is expected that 
the description of the role of electronic correlations in the 
unusual properties of the cuprate hight-temperature superconductors  
\cite{two-gaps,k-r-spaces,2D-MIT,ARPES-review} requires as well accounting
for the effects of three-dimensional uniaxial anisotropy
and intrinsic disorder.

The spin-singlet two-spinon $s1$ bond particle configurations studied
in this paper are an example of the general resonating-valence-bond pictures for spin-singlet 
occupancy configurations of ground states studied in Refs. \cite{Fazekas,Pauling}.
As mentioned above, the particular type of spin configurations based 
on two-spinon and two-site bonds considered in our studies lack both
the non-orthogonality and overcompleteness problems. 
Such spin-singlet two-spinon $s1$ bond particle configurations
refer to the $c$ and $s1$ fermion description
also used in the investigations of Ref. \cite{companion} on the Hubbard
model on the square lattice. Some progress in understanding the physics 
of that model has been achieved for different limits, through a variety 
of methods. It corresponds to a non-perturbative quantum problem in 
terms of electron operators, so that rewriting the theory in terms of the 
standard formalism of many-electron physics is an extremely complex problem. 
A detailed and extensive discussion of the relation between the new results 
obtained both in Ref. \cite{companion} and this paper by means
of the $c$ and $s1$ fermion description and previously known 
results is presented in that reference. 

For instance, it is shown that the predictions of the square-lattice quantum 
liquid theory concerning the spin spectrum at half filling as described in terms 
of the spin-singlet two-spinon configurations studied in this paper agree both 
with experiments on the parent compound La$_2$CuO$_4$ and results obtained by 
the standard formalism of many-body physics. Furthermore, in Ref. \cite{companion}
the relation of the description used in the studies of this paper
and that reference to several other schemes is discussed.

The paper is organized as follows. The model, the 
basic rotated-electron representation, and the description 
in terms of the basic quantum objects, which emerge
from the rotated-electron occupancy
configurations, are the subjects of Section 2.
The $s1$ effective lattice, corresponding change of 
gauge structure, and $s1$ operators of the $N^h_{s1}=0$ configuration state
are the topics addressed in Section 3.
(Here $N^h_{s1}$ denotes the number of unoccupied sites of the $s1$ 
effective lattice.) Section 4 contains
a study of the $N^h_{s1}=1,2$ configuration states and 
kink-like and anti-kink-like link occupancies associated with 
the $s1$ effective lattice unoccupied sites. Finally, Section 5 contains the 
concluding remarks.

\section{The Hubbard model on the square lattice, rotated electrons,
and the description used in our studies}

\subsection{The Hubbard model on the square lattice and rotated electrons}

The Hubbard model on the two-dimensional (2D) square lattice with 
torus periodic boundary conditions and the same model on the 1D
lattice with periodic boundary conditions, spacing $a$, 
$N_a^D\equiv [N_a]^D$ sites where $D=1$ and $D=2$ 
for the 1D and square lattices, respectively, $N_a^D\gg1$ even, 
and lattice edge length $L=N_a\,a$ 
for 2D and chain length $L=N_a\,a$ for 1D is given by,
\begin{equation}
\hat{H} = -t\sum_{\langle\vec{r}_j\vec{r}_{j'}\rangle}\sum_{\sigma =\uparrow
,\downarrow}[c_{\vec{r}_j,\sigma}^{\dag}\,c_{\vec{r}_{j'},\sigma}+h.c.] + 
U\,[N_a^D-\hat{Q}]/2 \, .
\label{H}
\end{equation}
Here the operator, 
\begin{equation}
{\hat{Q}} = \sum_{j=1}^{N_a^D}\sum_{\sigma =\uparrow
,\downarrow}\,n_{\vec{r}_j,\sigma}\,(1- n_{\vec{r}_j,-\sigma}) \, ,
\label{Q-op}
\end{equation}
where $n_{{\vec{r}}_j,\sigma} = c_{\vec{r}_j,\sigma}^{\dag} c_{\vec{r}_j,\sigma}$
and $-\sigma=\uparrow$ (and $-\sigma=\downarrow$)
for $\sigma =\downarrow$ (and $\sigma =\uparrow$)
counts the number of electron singly occupied sites.
Hence the operator ${\hat{D}}=[{\hat{N}}-{\hat{Q}}]/2$
counts the number of electron doubly
occupied sites where ${\hat{N}} = \sum_{\sigma}
{\hat{N}}_{\sigma}$ and ${\hat{N}}_{\sigma}=\sum_{j=1}^{N_a^D}
n_{{\vec{r}}_j,\sigma}$. 

The studies of Ref. \cite{companion} use a uniquely-defined
electron - rotated-electron unitary $\hat{V}=\hat{V}(U/4t)$. It is such that
states $\vert \Psi_{U/4t}\rangle =
{\hat{V}}^{\dag}\vert\Psi_{\infty}\rangle$ are for $U/4t>0$
energy eigenstates. It corresponds
to a suitable chosen set $\{\vert\Psi_{\infty}\rangle\}$
of $U/4t\rightarrow\infty$ energy eigenstates.
The unitary transformation maps the electronic operators
$c_{\vec{r}_j,\sigma}^{\dag}$ and $c_{\vec{r}_j,\sigma}$
onto rotated-electron creation and annihilation operators 
${\tilde{c}}_{\vec{r}_j,\sigma}^{\dag} =
{\hat{V}}^{\dag}\,c_{\vec{r}_j,\sigma}^{\dag}\,{\hat{V}}$ and 
${\tilde{c}}_{\vec{r}_j,\sigma} =
{\hat{V}}^{\dag}\,c_{\vec{r}_j,\sigma}\,{\hat{V}}$, respectively.

The studies of Ref. \cite{bipartite} reveal that for $U/4t>0$ 
the Hubbard model on a square lattice has a global $SO(3)\times SO(3)\times U(1)$ 
symmetry. The generator ${\tilde{S}}_c$ of the hidden global $U(1)$ symmetry 
reads ${\tilde{S}}_c= {\hat{V}}^{\dag}\,{\hat{S}}_c\,{\hat{V}}$
where ${\hat{S}}_c= {\hat{Q}}/2$ and the operator
${\hat{Q}}$ is given in Eq. (\ref{Q-op}). Its eigenvalue $S_c$
is one-half the number of rotated-electron singly occupied
sites $2S_c$. Indeed, the rotated electrons of Refs. \cite{companion,bipartite}
are constructed in such a way that $S_c$ is a good quantum number in the
rotated-electron picture. The subspaces spanned by states whose number 
$2S_c$ is constant play an important role. For hole concentrations $0\leq x<1$ and
maximum spin density $m=(1-x)$ there is a fully polarized  
vacuum, which remains invariant under the electron - rotated-electron unitary 
transformation,
\begin{equation}
\vert 0_{\eta s}\rangle = \vert 0_{\eta};N_{a_{\eta}}^D\rangle\times\vert 0_{s};N_{a_{s}}^D\rangle
\times\vert GS_c;2S_c\rangle \, .
\label{vacuum}
\end{equation}
Here the $\eta$-spin $SU(2)$ vacuum $\vert 0_{\eta};N_{a_{\eta}}^D\rangle$ 
associated with $N_{a_{\eta}}^D=[N_a^D-2S_c]$ independent $+1/2$
$\eta$-spinons, the spin $SU(2)$ vacuum $\vert 0_{s};N_{a_{s}}^D\rangle$ 
with $N_{a_{s}}^D=2S_c$ independent $+1/2$ spinons, and the $c$ $U(1)$
vacuum $\vert GS_c;2S_c\rangle$ with $N_c=2S_c$ $c$ fermions
remain invariant under the electron - rotated-electron unitary transformation.   
In turn, for states with a finite number of $s1$ bond particles the $c$ fermions
are not invariant under that transformation.

The whole physics can be extracted from the model (\ref{H}) in 
the LWS-subspace referring to values
of $S_{\alpha}$ and $S^z_{\alpha}$ such that 
$S_{\alpha}=-S^z_{\alpha}$ for $\alpha =\eta,s$ \cite{companion}. 
Within the LWS representation, the 
$c$ fermion creation operator can be expressed in terms
of the rotated-electron operators as follows,
\begin{equation}
f_{\vec{r}_j,c}^{\dag} =
{\tilde{c}}_{\vec{r}_j,\uparrow}^{\dag}\,
(1-{\tilde{n}}_{\vec{r}_j,\downarrow})
+ e^{i\vec{\pi}\cdot\vec{r}_j}\,{\tilde{c}}_{\vec{r}_j,\uparrow}\,
{\tilde{n}}_{\vec{r}_j,\downarrow} \, ,
\label{fc+}
\end{equation}
where ${\tilde{n}}_{\vec{r}_j,\sigma}=
{\tilde{c}}_{\vec{r}_j\sigma}^{\dag}{\tilde{c}}_{\vec{r}_j\sigma}$ and
$e^{i\vec{\pi}\cdot\vec{r}_j}$ is $\pm 1$ depending on which
sublattice site $\vec{r}_j$ is on. The three spinon local operators
$s^l_{\vec{r}_j}$ and three $\eta$-spinon local operators $p^l_{\vec{r}_j}$
such that $l=\pm,z$ and $s^{\pm}_{\vec{r}_j}= s^{x}_{\vec{r}_j}\pm i\,s^{y}_{\vec{r}_j}$
and $p^{\pm}_{\vec{r}_j}= p^{x}_{\vec{r}_j}\pm i\,p^{y}_{\vec{r}_j}$,
respectively, where the Cartesian coordinates $x,y,z$ are often denoted in this paper
by $x_1,x_2,x_3$, respectively, are given by,
\begin{equation}
s^l_{\vec{r}_j} = n_{\vec{r}_j,c}\,q^l_{\vec{r}_j} \, ; \hspace{0.50cm}
p^l_{\vec{r}_j} = (1-n_{\vec{r}_j,c})\,q^l_{\vec{r}_j} \, , 
\hspace{0.15cm} l =\pm,z \, ; \hspace{0.50cm}
n_{\vec{r}_j,c} = f_{\vec{r}_j,c}^{\dag}\,f_{\vec{r}_j,c} \, .
\label{sir-pir}
\end{equation}
Here $n_{\vec{r}_j,c} = f_{\vec{r}_j,c}^{\dag}\,f_{\vec{r}_j,c}$
is the $c$ fermion local density operator and the
rotated quasi-spin operators read as follows in terms of rotated-electron 
creation and annihilation operators,
\begin{equation}
q^+_{\vec{r}_j} = ({\tilde{c}}_{\vec{r}_j,\uparrow}^{\dag}
- e^{i\vec{\pi}\cdot\vec{r}_j}\,{\tilde{c}}_{\vec{r}_j,\uparrow})\,
{\tilde{c}}_{\vec{r}_j,\downarrow} \, ; \hspace{0.50cm}
q^-_{\vec{r}_j} = (q^+_{\vec{r}_j})^{\dag} \, ;
\hspace{0.50cm}
q^z_{\vec{r}_j} = {1\over 2} - {\tilde{n}}_{\vec{r}_j,\downarrow} \, .
\label{rotated-quasi-spin}
\end{equation}

Inversion of the relations provided in 
Eqs. (\ref{fc+}) and (\ref{rotated-quasi-spin}) gives,
\begin{eqnarray}
{\tilde{c}}_{\vec{r}_j,\uparrow}^{\dag} & = &
f_{\vec{r}_j,c}^{\dag}\,\left({1\over 2} +
q^z_{\vec{r}_j}\right) + e^{i\vec{\pi}\cdot\vec{r}_j}\,
f_{\vec{r}_j,c}\,\left({1\over 2} - q^z_{\vec{r}_j}\right) \, ,
\nonumber \\
{\tilde{c}}_{\vec{r}_j,\downarrow}^{\dag} & = &
q^-_{\vec{r}_j}\,(f_{\vec{r}_j,c}^{\dag} -
e^{i\vec{\pi}\cdot\vec{r}_j}\,f_{\vec{r}_j,c}) \, .
\label{c-up-c-down}
\end{eqnarray}

For the one- and two-electron subspace $N_a^D\gg 1$ and the number $N_{a_{s1}}^D$
of sites of the $s1$ effective lattice, $N_{s1}$ of $s1$ bond particles,
and $N^h_{s 1}$ of unoccupied sites read,
\begin{eqnarray}
N_{a_{s1}}^D & = & N_{s1} + N^h_{s1} = 
N_{a_s}^D/2 + S_s 
\, ; \hspace{0.35cm}
N_{s1} = N_{a_s}^D/2 -[S_s +2N_{s2}] \, ,
\nonumber \\
N^h_{s 1} & = & [2S_s +2N_{s2}] =0,1,2 \, ,
\label{Nas1-Nhs1}
\end{eqnarray}
whereas for the corresponding $c$ fermions such numbers are given by,
\begin{equation}
N_{a_{c}}^D = N_{c} + N^h_{c} = 
N_{a}^D 
\, ; \hspace{0.35cm}
N_{c} = 2S_c = (1-x)\,N_{a}^D  
\, ; \hspace{0.35cm}
N^h_{c} = x\,N_{a}^D \, .
\label{Nac-Nhc}
\end{equation}
For that subspace the $s1$ effective lattice is either full
with $N_{s1}=N_{a_{s1}}^D =N_{a_s}^D/2$ or
has one or two unoccupied sites. 
For it there is commensurability 
between the real-space distributions of the $N_{a_{s1}}^D\approx N_{s1}$
sites of the $s1$ effective lattice and $N_{a_{s}}^D\approx 2N_{s1}$ 
sites of the spin effective lattice. For $(1-x)\geq 1/N_a^D$ the spin effective lattice has 
$N_{a_s}^D=(1-x)\,N_a^D$ sites. The $N_{a_{s1}}^D$ expression given 
in Eq. (\ref{Nas1-Nhs1}) implies that $a_{s1} = L/N_{a_{s1}}$ reads,
\begin{equation}
a_{s1} = 2^{1/D}\,{a_s\over \left(1+{2S_s\over (1-x)N_a^D}\right)^{1/D}}
\approx 2^{1/D}\,a_s\, \left(1-{2S_s\over D(1-x)}{1\over N_a^D}\right)
\approx 2^{1/D}\,a_s \, ,
\label{a-a-s1-sube}
\end{equation}
where $2S_s =0, 1, 2$ and the lattice constant $a_s$ of the spin effective lattice is given in Eq. (\ref{NNCC}). 

For $N^h_{s1}=0$ states
the bipartite 1D and square spin effective lattices have two
well-defined sub-lattices. For the square lattice the two spin 
effective sub-lattices have lattice constant $a_{s1}=\sqrt{2}\,a_s$. 
In turn, for 1D the sites of each spin effective sub-lattice are 
distributed alternately along the chain, the corresponding 
nearest-neighboring sites being separated by $a_{s1} =2a_{s}$.
The fundamental translation vectors of such sub-lattices read,
\begin{equation}
{\vec{a}}_{s1} = a_{s1}\,{\vec{e}}_{x_1}
\, , \hspace{0.10cm} [1D] 
\, ; \hspace{0.35cm}
{\vec{a}}_{s1} = {a_{s1}\over\sqrt{2}}({\vec{e}}_{x_1}+{\vec{e}}_{x_2})
\, ; \hspace{0.35cm}
{\vec{b}}_{s1} = -{a_{s1}\over\sqrt{2}}({\vec{e}}_{x_1}-{\vec{e}}_{x_2}) 
\, , \hspace{0.10cm} [2D] \, ,
\label{a-b-s1}
\end{equation}
where ${\vec{e}}_{x_1}$ and ${\vec{e}}_{x_2}$ are the unit vectors. The vectors 
given in Eq. (\ref{a-b-s1}) are the fundamental translation vectors 
of the $s1$ effective lattice \cite{companion}.

The $c$ fermions are $\eta$-spinless and spinless fermions without internal structure and their effective lattice 
is identical to the original lattice. In contrast, the composite two-spinon $s1$ bond particles have internal structure
and expression of their $s1$ effective lattice occupancies in terms of spinon occupancies of 
the spin effective lattice and rotated-electron occupancies of the original lattice is
a more complex problem, which deserves and requires
further studies. This is the main subject of the remaining
of this paper.
    
\section{The $N^h_{s1}=0$ configuration state:
$s1$ effective lattice, corresponding change of 
gauge structure, and $s1$ bond-particle operators}

For the square-lattice model in the one- and two-electron subspace
the numbers $N_{s1}$, $N_{a_{s1}}^D$, and $N^h_{s1}=[N_{a_{s1}}^D-N_{s1}]$
are good quantum numbers whose values are fully controlled 
by those of $S_s^z$, $S_s$, and $S_c$.
We call {\it configuration states} the spinon occupancy configurations in the
spin effective lattice which
generate the spin degrees of freedom of the energy eigenstates that span
the one- and two-electron subspace. Such configuration states refer to
the overall occupancy of the $N_{s1}$ $s1$ bond particles over the
$N_{a_{s1}}^D$ sites of the $s1$ effective
lattice. Each configuration state
refers to well-defined positions of the $N^h_{s1}$
unoccupied sites. A $x\geq 0$ and $m=0$ ground
state is an example of an energy eigenstate whose spin
degrees of freedom are described by 
the $N^h_{s1}=0$ configuration state studied in the following.

We start by considering the $N^h_{s1}=[N_{a_{s1}}^D-N_{s1}]=0$ configuration state 
with no unoccupied sites in the spin and $s1$
effective lattices. It describes the spin degrees of freedom of
$x\geq 0$ and $m=0$ ground states and its charge excited states 
belonging to the one- and two-electron subspace whose $N_{a_{s1}}^D =
N_{a_s}^D/2=(1-x)\,N_a^D/2$ sites of the $s1$ effective lattice 
are occupied. Such a configuration state is 
described by suitable spinon occupancy configurations
of the $N_{a_s}^D=(1-x)\,N_a^D$ sites of the spin
effective lattice. 

\subsection{Independent two-site one-link bonds}

For simplicity we consider that 
$N_{a_s}=(1-x)^{1/D}\,N_a$ is an integer number 
yet if otherwise one reaches the same results in the 
$N_a^D\gg 1$ limit that our study refers to. For
the model on the square lattice the spin effective lattice is then a square lattice  
with $N_{a_s}\times N_{a_s}$ sites. For the model on
that lattice we consider torus periodic boundary conditions for the
spin effective lattice, alike for the original lattice. That 
implies periodic boundary conditions for the $N_{a_s}$
rows and $N_{a_s}$ columns. Periodic boundary conditions
are also used for the one-chain spin effective lattice
of the 1D model.

The spin effective lattice has in the present case two sub-lattices. The real-space 
coordinates of the sites of each of such sub-lattices correspond to 
a possible choice of those of the $s1$ effective lattice. Indeed, there is for the $N^h_{s1}=0$ configuration
state a gauge ``symmetry" between 
the representations in terms of the occupancies of the
two alternative choices of real-space coordinates of
the $s1$ effective lattice, which are found to refer
to two alternative and equivalent representations of that state. We 
use the same notation as Xiao-Gang Wen in 
Ref. \cite{Xiao-Gang} and say that there is a gauge structure when we use
simultaneously the real-space coordinates of the two 
corresponding choices of $s1$ effective 
lattices to label the $N^h_{s1}=0$ configuration
state. The real-space coordinates ${\vec{r}}_{j}$ of such two 
sub-lattices have $j=1,...,N_{a_{s1}}^D$ sites, in either case
the fundamental translation vectors being those given in 
Eq. (\ref{a-b-s1}). The real-space coordinates ${\vec{r}}_{j}$ of the $s1$ bond 
particles are chosen to correspond to those of one of these two
sub-lattices. Throughout the remaining of this paper
we call {\it sub-lattice 1} and {\it sub-lattice 2} the
sub-lattice of the spin effective lattice of a $N_{s1}^h=0$ configuration
state whose real-space
coordinates are and are not the same
as those of the $s1$ effective lattice,
respectively.  

An one-link bond connects two sites of the spin effective
lattice whose spinons have opposite spin projection and correspond to
a two-site spin-singlet configuration defined below.
Each $s1$ bond particle is a suitable superposition
of a well-defined set of two-site one-link bonds. Such a spin-singlet
two-spinon $s1$ bond particle is related to the
resonating-valence-bond pictures for spin-singlet 
occupancy configurations of ground states studied in Ref. \cite{Fazekas,Pauling}.
However, the $s1$ bond particle is well defined for all values of $U/4t>0$ 
and not only for $U/4t\gg 1$, consistently with its two spinons 
referring to spins of the sites singly occupied by rotated electrons. 
In contrast, most schemes used 
previously for the Hubbard model or related
models involving singly-occupied-site spins refer in general to 
large values of $U/4t\gg 1$ only \cite{2D-MIT,Feng,Fazekas,Xiao-Gang}.
Here the $s1$ bond particles have been constructed to inherently involving
spinon occupancy configurations of sites of the spin effective lattice, which for $U/4t>0$ refers
only to the sites of the original lattice singly occupied by 
rotated electrons.

For simplicity we often call  {\it two-site link} or just {\it link} a 
two-site one-link bond. For the $N^h_{s1}=0$ configuration state
studied here the above superposition includes $2D=2,4$ families of 
two-site links, each family having $N_{s1}/2D$ different types of
such links: $N_{s1}/2D$ is the largest number of independent 
links with the same link centre that exist for the 
above-considered boundary conditions. (Above and
in the remaining of this paper we denote often
the number of family links by $2D=2,4$ where
two and four is the number of such families for
the model on the 1D and square lattice, respectively.)
Link independence means here that for 
a given link centre all links involve 
different pairs of sites and each site belongs to one
pair only. 

The set of independent links with the same link centre belong to the same link family. 
Each one-link bond of a given family has some weight which in some cases may 
vanish. For a $s1$ bond particle of real-space coordinate ${\vec{r}}_{j}$ 
there are $2D=2,4$ families of links. The links of each family
are centered at one of the $2D=2,4$ points
${\vec{r}}_{j}+{\vec{r}_{d,l}}^{\,0}$ where the indexes $d=1,2$ 
for $D=2$, $d=1$ for $D=1$, and $l=\pm 1$
uniquely define the link family. Here ${\vec{r}_{d,l}}^{\,0}$ is the
primary link vector. It connects the site of real-space coordinate
${\vec{r}}_{j}$ to the centre of the four (and two for 1D) links
of real-space coordinate ${\vec{r}}_{j}+{\vec{r}_{d,l}}^{\,0}$ in the 
spin effective lattice 
involving that site and its nearest-neighboring sites. 
Note that the former site and the latter four
(and two for 1D) sites belong to sub-lattice 1 and
2, respectively. On choosing one of the 
two sub-lattices of the spin effective lattice to be
sub-lattice 1 and thus playing the role
of $s1$ effective lattice and representing the states
in terms of the occupancy configurations of the latter
lattice we say that there is a change of gauge structure 
\cite{Xiao-Gang}.
By considering all $N_{s1}/2D$ independent links for each 
of the $2D$ link centres needed to describe a $s1$ bond 
particle of real-space coordinate ${\vec{r}}_{j}$ we consider
the most general situation, the exact configuration referring
to some choice of the one-link bond weights considered below,
some of which may vanish.
  
A $s1$ bond particle of real-space coordinate ${\vec{r}}_{j}$
involves $N_{s1}$ two-site one-link bonds consistently with each family
having $N_{s1}/2D$ links of different type. The link type
is labeled by an index $g=0,1,...,[N_{s1}/2D-1]$
uniquely defined below.  
Each link of a $s1$ bond particle of real-space coordinate
${\vec{r}}_{j}$ involves two sites of coordinates 
$\vec{r}-\vec{r}_{d,l}^{\,g}$ and $\vec{r}+
\vec{r}_{d,l}^{\,g}$ where ${\vec{r}}_{j}=\vec{r}-\vec{r}_{d,l}^{\,0}$
so that the link centre $\vec{r}\equiv\vec{r}_{j}+{\vec{r}_{d,l}}^{\,0}$ is 
the middle point located half-way between the two sites
and the link vector $\vec{r}_{d,l}^{\,g}$ is defined below.
The real-space coordinates ${\vec{r}}_{j}=\vec{r}-\vec{r}_{d,l}^{\,g}$
and $\vec{r}+\vec{r}_{d,l}^{\,g}$ belong to the sub-lattice 1
and sub-lattice 2 of the spin effective lattice, respectively.
For each family there are $N_{s1}/2D$ link vectors
$\vec{r}_{d,l}^{\,g}$ which for the square lattice read,
\begin{equation}
\vec{r}_{d,l}^{\,g} = {\vec{r}_{d,l}}^{\,0}
+ {\vec{T}}_{d,l}^{\,g} \, ; \hspace{0.50cm}
{\vec{r}_{d,l}}^{\,0} = l\,{a_s\over 2}\,{\vec{e}}_{x_d}
\, ; \hspace{0.50cm} g=0,1,...,[N_{s1}/2D-1] \, ,
\label{r-r0-T}
\end{equation}
where $d=1,2$, $l = \pm 1$, and ${\vec{T}}_{d,l}^{\,g}$ is a $T$ vector. It has Cartesian
components ${\vec{T}}_{d,l}^{\,g}=[T_{d,l,1}^{\,g},T_{d,l,2}^{\,g}]$ 
for the square lattice and $\vec{T}_{1,l}^{\,g}=[T_{1,l,1}^{\,g}]$
for 1D. There 
are $N_{s1}/2D$ $T$ vectors ${\vec{T}}_{d,l}^{\,g}$,
one for each choice of the following Cartesian components,
\begin{eqnarray}
T_{d,l,i}^{\,g} & = & l\,a_s\,N_{i} 
\, ; \hspace{0.50cm} i = 1, 2 \, ,
\nonumber \\
N_d & = & 0,1,...,N_{a_s}/4 -1  
\, ; \hspace{0.15cm}
N_{\bar{d}} = -N_{a_s}/4 +1,...,-1,0,1,...,N_{a_s}/4 \, .
\label{xd-xd}
\end{eqnarray}
Here $d = 1,2$, $\bar{1} = 2$, $\bar{2} = 1$,
$ l = \pm 1$ and $N_d$ and $N_{\bar{d}}$ are consecutive integer 
numbers. The expressions provided in Eq. (\ref{r-r0-T}) apply to the
1D lattice as well provided that only the index value $d=1$
is considered. The 1D component of the $T$ 
vectors is given by $T_{1,l,1}^{\,g} = l\,a_s\,N_1$ where
$N_1 = 0,1,...,N_{a_s}/4 -1$. 

The link-type index $g=0,1,...,[N_{s1}/2D-1]$ labels 
the $N_{s1}/2D$ $T$ vectors 
${\vec{T}}_{d,l}^{\,g}$. For the square
lattice it is defined in terms of the numbers $N_d$
and $N_{\bar{d}}$ given in Eq. (\ref{xd-xd}) and reads, 
\begin{eqnarray}
g & = & N_d + 2\vert N_{\bar{d}}\vert {N_{a_s}\over 4} 
\, ; \hspace{0.50cm} N_{\bar{d}} \leq 0 \, ,
\nonumber \\
& = & N_d + 2(N_{\bar{d}}-1){N_{a_s}\over 4} 
\, ; \hspace{0.50cm} N_{\bar{d}} > 0 \, .
\label{g-2D}
\end{eqnarray}
For 1D one has that $g=N_1=0,1,...,N_{s1}/2 -1$.

The values of the link-type index $g$ are consecutive positive 
integers whose minimum value $g=0$ corresponds to
$N_1=N_2=0$ so that,
\begin{equation}
T_{d,l}^{\,0} = 0 \, .
\label{T0}
\end{equation}
For the model on the square lattice the maximum value 
is $g=[N_{s1}/2D-1]$ and refers to $N_d=N_{a_s}/4 -1$
and $N_{\bar{d}} =N_{a_s}/4$. 

Each pair of values (and value) of the Cartesian coordinates of 
the $T$ vector ${\vec{T}}_{d,l}^{\,g}=[T_{d,l,1}^{\,g},T_{d,l,2}^{\,g}]$ 
for the square lattice (and $\vec{T}_{1,l}^{\,g}=[T_{1,l,1}^{\,g}]$
for 1D) corresponds to exactly one of the values  
$g=0,1,...,[N_{s1}/2D-1]$ so that,
\begin{equation}
\sum_{N_d=0}^{N_{a_s}/4-1}
\sum_{N_{\bar{d}}=-N_{a_s}/4+1}^{N_{a_s}/4} \equiv 
\sum_{g=0}^{N_{s1}/4-1}  
\,  ; \hspace{0.50cm} D=2 
\, ; \hspace{0.50cm}
\sum_{N_1=0}^{N_{a_s}/4-1} \equiv  
\sum_{g=0}^{N_{s1}/2-1}  
\,  ; \hspace{0.50cm} D=1
\, .
\label{link-sum}
\end{equation}

Two-site links with the same $g$ value
and $d\neq d'$ and/or $l\neq l'$ are
{\it equivalent links}. Those are of the same
type but belong to different families.
Furthermore, $T$ vectors
${\vec{T}}_{d,l}^{\,g}$ and ${\vec{T}}_{d',l'}^{\,g}$
with the same value of $g$ and $d\neq d'$ and/or 
$l\neq l'$ are related as follows,
\begin{equation}
{\vec{T}}_{d,l}^{\,g} = ll'\left[\delta_{d,d'} +
\delta_{{\bar{d}},d'}{\bf \sigma_x}\right]
{\vec{T}}_{d',l'}^{\,g} \, ; \hspace{0.50cm}
{\bf \sigma_x} =\left[
\begin{array}{cc}
0 & 1 \\
1 & 0 \nonumber
\end{array}\right] \, ,
\label{T-T'-g}
\end{equation}
where ${\bf \sigma_x}$ is the usual Pauli matrix. 

As described below in terms of suitable
operators, a $s1$ bond particle of real-space coordinate
$\vec{r}_j$ is a superposition of $N_{s1}$
two-site one-link bonds each being associated with a 
link vector $\vec{r}_{d,l}^{\,g}$. For each site of the
spin effective lattice there is {\it exactly one}
other site of the same lattice such that the link connecting
the two sites has centre at $\vec{r}_{j}+{\vec{r}_{d,l}}^{\,0}$.
Therefore, any link of the same family involves two
sites of well-defined real-space coordinate 
$\vec{r}_{j}+{\vec{r}_{d,l}}^{\,0}-\vec{r}_{d,l}^{\,g}$ and 
$\vec{r}_{j}+{\vec{r}_{d,l}}^{\,0}+\vec{r}_{d,l}^{\,g}$,
which do not contribute together to any other link
of the same family. 

An important quantity is the distance between the two sites of
a link which we call {\it link length} or {\it two-site bond length}. It is independent of the real-space 
coordinate $\vec{r}=\vec{r}_{j}+{\vec{r}_{d,l}}^{\,0}$ of the link centre 
and is fully determined by the link vector $\vec{r}_{d,l}^{\,g}$ and
thus depends on the link type associated with the index $g$
only. For $D=2$ and $D=1$ it reads,
\begin{equation}
\xi_{g} \equiv \vert 2\vec{r}_{d,l}^{\,g}\vert  
= a_s\sqrt{(1+2N_d)^2 + (2N_{\bar{d}})^2} 
\, ; \hspace{0.5cm}
\xi_{g} \equiv \vert 2x_{1,l}^{\,g}\vert  
= a_s\,(1+2N_1) \, ,
\label{xi-L}
\end{equation}
respectively. Its minimum and maximum values are,
\begin{equation}
{\rm min}\,\xi_{g} = \xi_{0} = a_s 
\, ; \hspace{0.5cm}
{\rm max}\,\xi_{g} = \sqrt{2}\,a_s\,(N_{a_s}/2-1)
+ {\cal{O}} (1/N_a) \, ,
\label{max-min-xi-L}
\end{equation}
for the square lattice and 
${\rm min}\,\xi_{g} = \xi_{0} = a_s$
and ${\rm max}\,\xi_{g} = a_s\,(N_{a_s}/2-1)$
for 1D. 

For 1D, links with different $g$ have different
link length $\xi_{g}$. In turn, for
the square lattice there are links of different
type and hence different $g$ which have
the same link length $\xi_{g}$. Indeed,
analysis of the link-length expression
of Eq. (\ref{xi-L}) reveals that links with different $g$
value and numbers 
$[N_d,N_{\bar{d}}]$ and $[N_d',N_{\bar{d}}']$, 
respectively, such that $N_d=N_d'$ and 
$N_{\bar{d}}=-N_{\bar{d}}'$ have the same
link length. 

The set of values of the numbers $N_1$ and $N_2$
given in Eq. (\ref{xd-xd}) imply that the maximum value
of the link length is $\sqrt{D}\,a_s\,(N_{a_s}/2-1)$
rather than $\sqrt{D}\,a_s\,(N_{a_s}-1)$. Indeed and as
mentioned above, the links contributing 
to a $s1$ bond particle of the $N^h_{s1}=0$ configuration state 
are independent. It is then required that each
link involves two sites that participate simultaneously in 
exactly one of such links. Within the torus row and column periodic boundary
conditions for the square spin effective lattice 
such a requirement is fulfilled provided that 
the range of the numbers $N_1$ and $N_2$
is that given in Eq. (\ref{xd-xd}).  

For each family of two-site links associated with
a $s1$ bond particle of real-space coordinate $\vec{r}_{j}$
there is a {\it primary link}. It corresponds to $g=0$ and
thus connects two nearest-neighboring sites of the spin effective lattice,
one of them having the same real-space coordinate 
$\vec{r}_{j}=\vec{r}-{\vec{r}_{d,l}}^{\,0}$ as the $s1$ 
bond particle. For primary links the link vector $\vec{r}_{d,l}^{\,g}$
reads $\vec{r}_{d,l}^{\,g}={\vec{r}_{d,l}}^{\,0}$ where the primary
link vector ${\vec{r}_{d,l}}^{\,0}$ is
given in Eq. (\ref{r-r0-T}). 

For the model on the square lattice
there are four primary links, one per family.
Their link vectors ${\vec{r}_{d,l}}^{\,0}$ have
components such that $N_1=N_2=0$ in Eqs. (\ref{r-r0-T}) 
and (\ref{xd-xd}). Therefore, the primary links have minimum
length $\xi_{{\vec{r}}_{d,l}^{\,0}}=a_s$. Alike the 
remaining links of its family, the centre of a primary 
link is located at $\vec{r}=\vec{r}_{j}+{\vec{r}_{d,l}}^{\,0}$.
For the square lattice there are two horizontal primary links 
whose centers are located at $\vec{r}_{j}+\vec{r}_{1,l}^{\,0}$ with 
$l=\pm 1$ and two vertical primary links whose centers are
located at $\vec{r}_{j}+\vec{r}_{2,l}^{\,0}$ with $l=\pm 1$.
In the case of the 1D lattice there are two 
primary links whose centers are located at 
$\vec{r}_{j}+\vec{r}_{1,l}^{\,0}$ with $l=\pm 1$.

\subsection{Partitions and $g$-primary partitions}

The building blocks of the $N_{s1}^h=0$ configuration state
are singlet pairs of spinons on sites $\vec{r}_j^{\,-}$
and $\vec{r}_j^{\,+}$ of the spin effective lattice,
\begin{eqnarray}
\vert\vec{r}_j^{\,-},\vec{r}_j^{\,+}\rangle
& = & {1\over \sqrt{2}}\left(\vert\uparrow_{\vec{r}_j^{\,-}}
\downarrow_{\vec{r}_j^{\,+}}\rangle -
\vert\downarrow_{\vec{r}_j^{\,-}}
\uparrow_{\vec{r}_j^{\,+}}\rangle\right) \, ,
\nonumber \\
\vec{r}_j^{\,\mp} & = & \vec{r}_j + \vec{r}_{d,l}^{\,0}
\mp \vec{r}_{d,l}^{\,g} 
\, ; \hspace{0.5cm}
d = d (j) \, , \hspace{0.15cm}
l = l (j) \, , \hspace{0.15cm}
g = g (j) \, , 
\label{state-+}
\end{eqnarray}
where the values of the integer indexes $d$, $l$, and $g$ are
in the ranges $d=1,2$, $l=\pm 1$, and $g\in (0,N_{s1}/2D-1)$,
respectively, and are a function of the index $j=1,...,N_{s1}$ of the real-space coordinate
$\vec{r}_j$ in the sublattice 1 of each $s1$ bond particle.
Indeed, the two sites of such pairs of sites are connected by one-link bonds 
and each bond is associated with exactly one $s1$ bond particle. 

Each connection involving $N_{s1}$ different bonds determines a {\it partition}. 
A partition is a $2N_{s1}$-spinon occupancy configuration where each site of
the spin effective lattice is linked to one site only
and all $2N_{s1}$ sites then correspond to $N_{s1}$
well-defined two-site one-link bonds, each belonging to a different $s1$ bond 
particle. 

We recall that a $s1$ bond particle involves the superposition of 
$N_{s1}$ such two-site one-link bonds whereas a partition involves one 
two-site one-link bond from each of the $N_{s1}$ $s1$ bond particles.
Each $s1$ bond particle contributes with exactly one of its 
two-site bonds to a partition. In a partition any site of the spin effective 
lattice participates in one bond only and there is
a single link attached to each site which connects it to some
other site. And the latter site is attached to the former site only.
  
The $N_{s1}^h=0$ configuration state is then represented as,
\begin{equation}
\vert \phi \rangle =
\sum_{P} C_P
\prod_{j=1}^{N_{s1}}\vert\vec{r}_j^{\,-},\vec{r}_j^{\,+}\rangle 
\, ; \hspace{0.5cm}
C_P = \prod_{j=1}^{N_{s1}} h_{g(j)}^* \, ,
\label{phi-+}
\end{equation}
where the coefficients $h_g$ are associated with the bond
weights and appear in the bond operators 
defined below, the product of singlet states
$\prod_{j=1}^{N_{s1}}\vert\vec{r}_j^{\,-},\vec{r}_j^{\,+}\rangle$ 
refers to a bond state associated with a given partition, and
the summation $\sum_{P}$ is over all partitions.

A particular type of partition involves $N_{s1}$ 
identical links. The indexes $d$, $l$, and $g$ of identical links 
have the same values but correspond to $s1$ bond
particles with different real-space coordinates $\vec{r}_j$.
Such a partition involves a set of $N_{s1}$ identical 
two-site one-link bonds whose links connect different sites of the 
spin effective lattice, each site being linked to exactly one site.  
In this case the two real-space coordinates of the 
$N_{s1}$ pairs of sites are connected by the same 
real-space vector $2\vec{r}_{d,l}^{\,g}$ so that each bond 
link has the same length. 
\begin{figure}
\includegraphics[width=3.5cm,height=3.5cm]{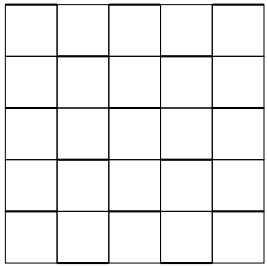}
\caption{Sub-domain of the spin effective lattice 
with a primary partition  of 
$d=1$ horizontal links for the model on the square lattice. The primary links are 
represented by the thick horizontal lines connecting 
two sites of the spin effective lattice. For a reference frame 
where the site located at the corner on the left-hand
side and lower limit of the squared sub-domain has 
Cartesian coordinates $(0,0)$ the family
indices read $d=1$, $l=+1$ if that site 
belongs to sub-lattice 2, whereas 
$d=1$, $l=-1$ if instead it belongs to 
sub-lattice 1.}
\end{figure}

Each of the $N_{s1}$ bonds of a partition 
involves two sites of real-space coordinates $\vec{r}_{j'}$ 
and $\vec{r}_{j'}+2\vec{r}_{d,l}^{\,g}$ which belong 
to different sub-lattices where $j'=1,...,N_{s1}$. The relation to the notation
used above for the real-space coordinates of the
two sites of a link is as follows,
\begin{equation}
\vec{r}_{j'} = \vec{r}_{j}+{\vec{r}_{d,l}}^{\,0}-\vec{r}_{d,l}^{\,g} 
\, ; \hspace{0.5cm}
\vec{r}_{j'}+2\vec{r}_{d,l}^{\,g} = 
\vec{r}_{j}+{\vec{r}_{d,l}}^{\,0}+\vec{r}_{d,l}^{\,g} \, ;
\hspace{0.15cm} j,j'=1,...,N_{s1} \, ,
\label{corresp}
\end{equation}
where both $\vec{r}_{j'}$ and $\vec{r}_{j}$ are real-space
coordinates of the sub-lattice 1 and thus of the $s1$ effective lattice. 
Except for a primary link one has that $j\neq j'$. The site of 
real-space coordinate $\vec{r}_{j}$ 
is that of the corresponding $s1$ bond
particle. It is closest to the link centre
at $\vec{r}_{j}+{\vec{r}_{d,l}}^{\,0}$. In turn,
$\vec{r}_{j'}$ is the real-space coordinate of one of
the two sites of the spin effective lattice involved
in the link. 
 
When a partition is a set of $N_{s1}$ identical primary
links, all site pairs involve nearest-neighboring sites of the 
spin effective lattice. It is then called a {\it primary partition}. 
The family of a primary partition is labeled by the indexes $d$
and $l$ of the corresponding identical links. Figures 1 and 2
represent primary partitions
of $d=1$ horizontal and $d=2$ vertical links, respectively, 
for a sub-domain of the spin effective lattice of the model on the 
square lattice.
\begin{figure}
\includegraphics[angle=90,width=3.5cm,height=3.5cm]{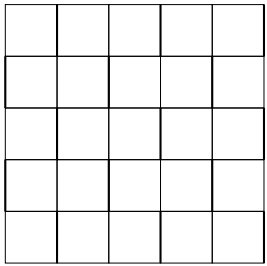}
\caption{Sub-domain of the spin effective lattice 
with a primary partition  of 
$d=2$ vertical links for the model on the square lattice. The primary links are 
represented by the thick vertical lines connecting 
two sites of the spin effective lattice. As in Fig. 1, 
the family indices read $d=2$, $l=+1$ if the site of
Cartesian coordinates $(0,0)$
belongs to sub-lattice 2, whereas 
$d=2$, $l=-1$ if instead it belongs to 
sub-lattice 1.}
\end{figure}

An useful concept is that of a {\it $g$-primary partition}. It is defined 
as the superposition of the $2D=2,4$ primary partitions. It follows 
that a $g$-primary partition contains $2D\,N_{s1}$ 
two-site primary links. In such a 
configuration each of the $2N_{s1}$ sites of the 
spin-effective lattice has $2D=2,4$ links attached to it.
Figure 3 shows a sub-domain of the spin-effective 
lattice with the $g$-primary partition 
of the $N^h_{s1}=0$ configuration state
for the model on the square lattice. 

\subsection{The $s1$ bond-particle operators and the subspace
they act onto}

The spinon occupancy configurations considered
above are similar to those associated
with multi-spin wave functions of spin-singlet states
used by several authors \cite{Fazekas,Fradkin,Auerbach}.
Such wave functions are often constructed having as 
building blocks two-site and two-spin spin-singlet configurations
similar to that of Eq. (\ref{state-+}) except
that here the two spins refer to sites singly
occupied by rotated electrons and thus correspond to
$U/4t>0$ rather than only to $U/4t\gg 1$. In such schemes
one also connects pairs of lattice sites with
bonds and each such a connection determines
a partition. However, here bonds involve sites of the
spin effective lattice whereas those of previous related
studies refer to the sites of the original lattice of
the corresponding quantum problems.
\begin{figure}
\includegraphics[width=3.75cm,height=4.50cm]{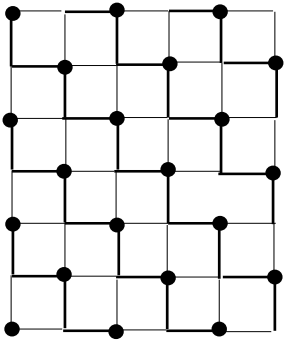}
\caption{Sub-domain of the spin-effective lattice 
representing the $g$-primary partition 
of the $N^h_{s1}=0$ configuration state
for the model on the square lattice. The sites belonging to the 
sub-lattice 1 are represented by 
filled circles. The horizontal (and vertical) thin 
and thick lines refer to $d=1$ (and $d=2$) $l=+1$ and 
$l=-1$ primary links, respectively. For the square lattice each 
site of the spin effective lattice has in a $g$-primary partition four links attached to it.}
\end{figure}

For a given partition one can define a valence bond state
\cite{Fazekas,Fradkin,Auerbach}
as a product of singlet states and represent
an arbitrary singlet by a superposition
of valence bond states of general form similar
to that of Eq. (\ref{phi-+}). That involves a sum
over all partitions of the lattice into set of pairs.
However, for general wave functions 
such a decomposition works in general very
badly. Indeed, valence-bond states are not
orthogonal and their basis is overcomplete.
Fortunately, here each
of the $N_{s1}$ two-site bonds of a partition belongs
to a different $s1$ bond particle so that
each of such particles contributes to a
partition with exactly
one bond. Such a restriction
eliminates the unwanted and unphysical
contributions and renders the bond states
free of the overcompleteness problem.

For the one- and two-electron subspace and $N_a\rightarrow\infty$, 
the operators $g_{{\vec{r}}_{j},s1}$ (and  $g^{\dag}_{{\vec{r}}_{j},s1}$)
which annihilate (and create) a 
$s1$ bond particle at a site of the spin effective lattice
of real-space coordinate ${\vec{r}}_{j}$ have the following 
general form both for the $1D$ and square lattices,
\begin{eqnarray}
g_{\vec{r}_{j},s1} & = & \sum_{g=0}^{N_{s1}/2D-1} h_{g}\, a_{\vec{r}_{j},s1,g}  
\, ; \hspace{0.15cm} g_{\vec{r}_{j},s1}^{\dag} = 
\left(g_{{\vec{r}}_{j},s1}\right)^{\dag} \, ,
\nonumber \\
a_{\vec{r}_{j},s1,g} & = &
\sum_{d=1}^{D}\sum_{l=\pm1}
\, b_{\vec{r}_{j}+{\vec{r}_{d,l}}^{\,0},s1,d,l,g} 
\, ; \hspace{0.50cm} D=1,2 \, ,
\label{g-s1+general}
\end{eqnarray}
so that the expression of $g_{\vec{r}_{j},s1}^{\dag} $ involves
the operators $a_{\vec{r}_{j},s1,g}^{\dag} = \left(a_{{\vec{r}}_{j},s1,g}\right)^{\dag}$
and $b_{\vec{r},s1,d,l,g}^{\dag} = \left(b_{\vec{r},s1,d,l,g}\right)^{\dag}$.
The operators $a_{\vec{r}_{j},s1,g}^{\dag}$ and $a_{\vec{r}_{j},s1,g}$ 
create and annihilate, respectively, a superposition of $2D=2,4$ bonds of 
the same type and $b_{\vec{r},s1,d,l,g}^{\dag}$ and $b_{\vec{r},s1,d,l,g}$ are
two-site one-bond operators whose expression is given below. 

For the square (and 1D) lattice the four (and two) primary links
associated with the operators $a_{\vec{r}_{j},s1,0}^{\dag}$ and 
$a_{\vec{r}_{j},s1,0}$ are behind most of the spectral weight 
of a $s1$ bond particle of real-space coordinate 
$\vec{r}_{j}$. Consistently, the absolute value $\vert h_{g}\vert$
of the coefficients $h_{g}$ appearing in the expressions
of such operators given in Eq. (\ref{g-s1+general}) 
decreases for increasing link length $\xi_{g}$. 
These coefficients obey the normalization sum-rule,
\begin{equation}
\sum_{g=0}^{[N_{s1}/2D-1]} \vert h_{g}\vert^2 = {1\over 2D} 
\,  ; \hspace{0.50cm} D=1,2 \, .
\label{g-s1+sum-rule}
\end{equation}
The exact dependence of $\vert h_{g}\vert$ on the
link length $\xi_{g}$, value of $U/4t$, and hole concentration
$x$ remains for the Hubbard model an involved open problem. The suitable use
of this sum-rule and related symmetries leads though to useful information,
as discussed below. By definition the real-space coordinates ${\vec{r}}_{j}$ of the
$s1$ bond-particle operators of Eq. (\ref{g-s1+general}) 
are those of the sub-lattice 1. 

The two-site one-bond operators $b_{\vec{r},s1,d,l,g}^{\dag}$ 
and $b_{\vec{r},s1,d,l,g}$
appearing in Eq. (\ref{g-s1+general}) are
associated with a well-defined link connecting the
two sites of real-space coordinates $\vec{r}-\vec{r}_{d,l}$
and $\vec{r}+\vec{r}_{d,l}$, respectively.
Their expression can be obtained by 
considering the following related operator, 
\begin{eqnarray}
& & {(-1)^{d-1}\over\sqrt{2}}
[(1-{\tilde{n}}_{\vec{r}-\vec{r}_{d,l}^{\,g},\downarrow})
{\tilde{c}}^{\dag}_{\vec{r}-\vec{r}_{d,l}^{\,g},\uparrow}\,
{\tilde{c}}^{\dag}_{\vec{r}+\vec{r}_{d,l}^{\,g},\downarrow}\, 
(1-{\tilde{n}}_{\vec{r}+\vec{r}_{d,l}^{\,g},\uparrow})
\nonumber \\
& - & (1-{\tilde{n}}_{\vec{r}-\vec{r}_{d,l}^{\,g},\uparrow})\,
{\tilde{c}}^{\dag}_{\vec{r}-\vec{r}_{d,l}^{\,g},\downarrow}\,
{\tilde{c}}^{\dag}_{\vec{r}+\vec{r}_{d,l}^{\,g},\uparrow}\,
(1-{\tilde{n}}_{\vec{r}+\vec{r}_{d,l}^{\,g},\downarrow})]
\nonumber \\
& = & f_{\vec{r}-\vec{r}_{d,l}^{\,g},c}^{\dag}\,f_{\vec{r}+\vec{r}_{d,l}^{\,g},c}^{\dag} \,
b_{\vec{r},s1,d,l,g}^{\dag} \, ,
\label{singlet-conf}
\end{eqnarray}
where in the second expression the operator 
$b_{\vec{r},s1,d,l,g}^{\dag}$ reads,
\begin{equation}
b_{\vec{r},s1,d,l,g}^{\dag} = 
{(-1)^{d-1}\over\sqrt{2}}\left(\left[{1\over 2}+s^z_{\vec{r}-\vec{r}_{d,l}^{\,g}}\right]
s^-_{\vec{r}+\vec{r}_{d,l}^{\,g}} - \left[{1\over 2}+s^z_{\vec{r}+\vec{r}_{d,l}^{\,g}}\right]
s^-_{\vec{r}-\vec{r}_{d,l}^{\,g}}\right) \, ,
\label{g-s-l}
\end{equation}
and $b_{\vec{r},s1,d,l,g} = \left(b_{\vec{r},s1,d,l,g}^{\dag}\right)^{\dag}$.
Here the spinon operators are those given in Eq. (\ref{sir-pir}).
The second expression of Eq. (\ref{singlet-conf}) is obtained by the use of 
Eq. (\ref{c-up-c-down}). 

The phase factor $(-1)^{d-1}$ which appears in the
operator of Eq. (\ref{g-s-l}) is associated with the $d$-wave symmetry of the
$s1$ bond-particle two-spinon pairing of the model on the square lattice.
The introduction of such a phase-factor refers to a self-consistent
procedure which follows from the $d$-wave symmetry of
the spinon energy dispersion found in Ref. \cite{companion} for
the $s1$ fermions. Such objects emerge from the
$s1$ bond particles studied here through a suitable extended Jordan-Wigner
transformation and their energy-dispersion $d$-wave symmetry
arises naturally from symmetries beyond the form of
the operators introduced in Eqs.
(\ref{singlet-conf}) and (\ref{g-s-l}).

According to the above discussions, the 
real-space coordinates $\vec{r}-\vec{r}_{d,l}^{\,g}$ and $\vec{r}+\vec{r}_{d,l}^{\,g}$ 
involved in the operators of Eqs. (\ref{singlet-conf}) and (\ref{g-s-l}) correspond to
two sites that belong to different sub-lattices of the 
spin effective lattice, $\vec{r}=\vec{r}_{j}+{\vec{r}_{d,l}}^{\,0}$ 
is the link centre, and
the primary link vector ${\vec{r}_{d,l}}^{\,0}$ and link vector
$\vec{r}_{d,l}^{\,g}$ are given in Eqs. (\ref{r-r0-T}) and (\ref{xd-xd}).
In the configuration generated by the operator of Eq. (\ref{singlet-conf})
the two sites are singly-occupied  by the rotated electrons 
associated with the operators appearing in the first expression 
of that equation. 

The $N^h_{s1}=0$ configuration state (\ref{phi-+}) can
be written in terms of $s1$ bond-particle operators
given in Eq. (\ref{g-s1+general}) as follows,
\begin{equation}
\vert \phi \rangle = \prod_{j=1}^{N_{s1}} g_{\vec{r}_{j},s1}^{\dag}\vert 0_{s};N_{a_{s}}^D\rangle \, ,
\label{phi-+-s1}
\end{equation}
where $\vert 0_{s};N_{a_{s}}^D\rangle$ is the spin $SU(2)$ vacuum 
with $N_{a_{s}}^D=2N_{s1}$ independent $+1/2$ spinons 
on the right-hand side of Eq. (\ref{vacuum}). It corresponds 
to a fully polarized spin-up configuration. 
The subspace where the operators of Eqs. (\ref{g-s1+general})-(\ref{g-s-l}) 
act onto is defined by imposing the equality of the general
configuration states given in Eqs. (\ref{phi-+}) and  (\ref{phi-+-s1}),
respectively. Such an equality implies several restrictions on the 
transitions generated by the two-site one-bond operators (\ref{g-s-l})
which are summarized in four corresponding rules for exclusion of unphysical
and unwanted spin configurations given below. Before introducing
such rules let us discuss several properties of the $s1$ bond-particle
operators which follow from the algebra given in Eqs. 
(\ref{albegra-s-p-m})-(\ref{albegra-s-sz-com}) of Appendix A
for the basic spinon operators of the two-site one-bond
operators expressions provided in Eq. (\ref{g-s-l}).

The $s1$ bond-particle operators $g^{\dag}_{{\vec{r}}_{j},s1}$
and $g_{{\vec{r}}_{j},s1}$ of Eq. (\ref{g-s1+general}) 
involve a sum of $N_{s1}$ two-site one-bond operators of general form
given in Eq. (\ref{g-s-l}) with $N_{s1}/2D$ 
of such operators per family.
The number of unoccupied sites $N^h_{s 1}$ of 
Eq. (\ref{Nas1-Nhs1})
refers to a subspace with constant number $N_{s1}$ of $s1$
bond particles. In turn, the creation and annihilation of one $s1$ bond particle by application
of these operators, respectively,
onto the ground state involves a superposition of
$N_{s1}$ elementary processes, which do not
conserve the number of these objects. Each 
such an elementary process is generated by 
an operator $b_{\vec{r},s1,d,l,g}^{\dag}$ and
$b_{\vec{r},s1,d,l,g}$, respectively, whose expression
is given in Eq. (\ref{g-s-l}). 

Within the present LWS representation, application
of the rotated-electron operators of 
Eq. (\ref{singlet-conf}) onto two unoccupied sites
of the original lattice generates two virtual processes. The first process
involves creation of two $c$ fermions and two 
independent $+1/2$ spinons. The second process
refers to creation of a spin-singlet two-site and two-spinon
configuration upon annihilation of two independent $+1/2$ spinons.
Indeed, from analysis of the expression provided in Eq. (\ref{g-s-l}), 
one finds that application of the operator $b_{\vec{r},s1,d,l,g}^{\dag}$  
onto the sites of real-space coordinates 
$\vec{r}-\vec{r}_{d,l}^{\,g}$ and $\vec{r}+
\vec{r}_{d,l}^{\,g}$ gives zero except when
those sites are both occupied by an independent
$+1/2$ spinon. This is consistent with a "unoccupied site" 
referring to two sites of the spin effective lattice with real-space 
coordinates $\vec{r}-\vec{r}_{d,l}^{\,g}$ and 
$\vec{r}+\vec{r}_{d,l}^{\,g}$, respectively, which are occupied by 
independent $+1/2$ spinons in the initial configuration 
state. 

As confirmed in Section 4 and consistently with the studies
of Ref. \cite{companion}, for excitations involving transitions
between configuration states where a $s1$ bond
particle moves around in the spin effective lattice
by elementary processes which conserve the
numbers $N_{s1}$ and $N^h_{s1}$, an independent 
$+1/2$ spinon plays the role of an unoccupied site 
of the $s1$ and spin effective lattices. In contrast, in 
elementary processes involving the creation of a $s1$ 
bond particle the two annihilated independent $+1/2$
spinons play the role of a "unoccupied site", which
becomes occupied in the final occupancy configuration. 
\begin{figure}
\includegraphics[width=5.04cm,height=1.8cm]{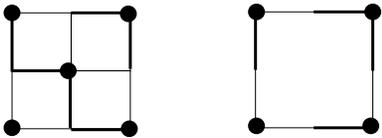}
\caption{Small sub-domain of two $g$-primary partitions
associated with $N^h_{s1}=0$ and $N^h_{s1}=2$ configuration states
for the model on the square lattice.
In the configuration of the figure on the left-hand side the 
filled circle at the middle of the sub-domain corresponds 
to the site whose real-space coordinate $\vec{r}_j$ is that of 
the $s1$ bond particle. Annihilation of that 
object leads to the configuration on the
figure right-hand side. Note that for a primary $g$-basic 
partition this is equivalent to the suppression of 
the four links attached to the above site of real-space coordinate 
$\vec{r}_j$.}
\end{figure}

According to the operator expression provided in Eq. (\ref{g-s-l}), 
upon acting onto the independent-spinon occupancies the operator 
$b_{\vec{r},s1,d,l,g}^{\dag}$ generates a superposition of two
configurations. For one of these configurations the elementary
process generated by that operator flips the spin of the spinon at 
site $\vec{r}+\vec{r}_{d,l}^{\,g}$ and checks that the spin 
of the spinon at site $\vec{r}-\vec{r}_{d,l}^{\,g}$ remains up. The 
elementary process generating the other configuration flips the spin
of the spinon at site $\vec{r}-\vec{r}_{d,l}^{\,g}$ and checks that the 
spin of the spinon at site $\vec{r}+\vec{r}_{d,l}^{\,g}$ remains up. 
The relative phase 
factor $-1$ of the two configurations insures that the
$s1$ bond particle created by the operator 
$g_{\vec{r}_j,s1}^{\dag}$ of Eq. (\ref{g-s1+general}) is a
suitable superposition of spin-singlet configurations. 

Figure 4 shows a small sub-domain of two
$g$-primary partitions of $N^h_{s1}=0$
and $N^h_{s1}=2$ configuration states.
In the configuration on the figure left-hand
side the filled circle at the middle of the sub-domain 
corresponds to a site whose real-space coordinate 
$\vec{r}_j$ is that of a $s1$ bond particle. Application 
of the annihilation operator $g_{\vec{r}_{j},s1}$ of Eq. (\ref{g-s1+general}) 
onto that $g$-primary partition occurs
through the operator $a_{\vec{r}_{j},s1,0}$ 
also given in that equation. That leads to the configuration
on the figure right-hand side. Hence
annihilation of the $s1$ bond particle is 
for a $g$-primary partition equivalent
to the suppression of the four links attached 
to the above site of real-space coordinate $\vec{r}_j$. 

A superficial analysis of the configurations shown
in Fig. 4 seems to indicate that there is one unoccupied 
site in the sub-domain of the final $N^h_{s1}=2$ configuration state. However, if
instead of the $g$-primary partition one
considers the corresponding four primary partitions one finds that there are two
nearest-neighboring unoccupied sites. For the 
$d=1$ and $d=2$ primary partitions
these two sites belong to the same row and column, 
respectively, as further discussed in Section 4.

Concerning the application onto spin configurations of 
two-site one-bond operators and four-site two-bond operators,
the restrictions arising from imposing that the representations
of the $N_{s1}^h=0$ configuration state given in 
Eqs. (\ref{phi-+}) and  (\ref{phi-+-s1}), respectively, 
are identical correspond to the following four rules whose fulfillment 
prevents the generation of unwanted and unphysical spin 
configurations:
\vspace{0.25cm}

{\it First rule} according to which application onto
a spin configuration of a two-site one-bond operator 
$b_{\vec{r},s1,d,l,g}^{\dag}$ or $b_{\vec{r},s1,d,l,g}$
generates a physical spin configuration provided that its 
sites of real-space coordinates $\vec{r}-\vec{r}_{d,l}^{\,g}$ and 
$\vec{r}+\vec{r}_{d,l}^{\,g}$, respectively, are in the initial spin configuration
(i) occupied by independent $+1/2$ spinons or (ii) linked by a bond. 
\vspace{0.25cm}

{\it Second rule} states that application onto a spin configuration of any of 
the elementary four-site two-bond operators 
$b_{\vec{r},s1,d,l,g}^{\dag}b_{\vec{r}',s1,d',l',g'}^{\dag}$,
$b_{\vec{r},s1,d,l,g}b_{\vec{r}',s1,d',l',g'}^{\dag}$,
$b_{\vec{r},s1,d,l,g}b_{\vec{r}',s1,d',l',g'}$, and
$b_{\vec{r},s1,d,l,g}^{\dag}b_{\vec{r}',s1,d',l',g'}$
gives zero when one of the two
sites of real-space coordinates $\vec{r}-\vec{r}_{d,l}^{\,g}$ 
and $\vec{r}+\vec{r}_{d,l}^{\,g}$, respectively,
is the same as one of the two
sites of real-space coordinates $\vec{r}'-\vec{r}_{d',l'}^{\,g'}$ 
and $\vec{r}'+\vec{r}_{d',l'}^{\,g'}$. Hence only
when the two two-site one-bond operators of such four-site
operators do not join sites or join both sites 
their application onto a spin configuration generates
physical spin configurations.
\vspace{0.25cm}

{\it Third rule} according to which when the two two-site 
one-bond operators of an elementary four-site two-bond operator
$b_{\vec{r},s1,d,l,g}^{\dag}b_{\vec{r}',s1,d',l',g'}^{\dag}$,
$b_{\vec{r},s1,d,l,g}b_{\vec{r}',s1,d',l',g'}^{\dag}$,
$b_{\vec{r},s1,d,l,g}b_{\vec{r}',s1,d',l',g'}$, and
$b_{\vec{r},s1,d,l,g}^{\dag}b_{\vec{r}',s1,d',l',g'}$,
correspond (i) to different $s1$ bond particles so that
in the real-space coordinates $\vec{r} = \vec{r}_j+\vec{r}_{d,l}^{\,0}$
and $\vec{r}' = \vec{r}_{j'}+\vec{r}_{d',l'}^{\,0}$
of their link centres, respectively,
one has that $j=j'$ or (ii) to the same 
$s1$ bond particle and to the same sites
so that $\vec{r}\pm\vec{r}_{d,l}^{\,g}=\vec{r}'\pm\vec{r}_{d',l'}^{\,g'}$  
the first rule applies independently to each of such two-site one-bond operators
provided that in case (i) the second rule is obeyed. 
\vspace{0.25cm}

{\it Forth rule} refers to when the real-space coordinates of the link centres
of the two-site one-bond operators of any of the four elementary four-site 
two-bond operators considered in the third rule 
are given by $\vec{r} = \vec{r}_j+\vec{r}_{d,l}^{\,0}$
and $\vec{r}' = \vec{r}_j+\vec{r}_{d',l'}^{\,0}$, respectively, 
so that such two-site one-bond operators correspond to the same 
$s1$ bond particle of real-space coordinate $\vec{r}_j$ but 
$\vec{r}_{d,l}^{\,g}\neq \vec{r}_{d',l'}^{\,g'}$ and states
that then there are two cases. When in the initial spin
configuration where the operators 
$b_{\vec{r},s1,d,l,g}^{\dag}b_{\vec{r}',s1,d',l',g'}^{\dag}$ and
$b_{\vec{r},s1,d,l,g}b_{\vec{r}',s1,d',l',g'}^{\dag}$
(and $b_{\vec{r},s1,d,l,g}b_{\vec{r}',s1,d',l',g'}$ and
$b_{\vec{r},s1,d,l,g}^{\dag}b_{\vec{r}',s1,d',l',g'}$)
act onto the sites of the spin effective lattice of real-space 
coordinates $\vec{r}_j+\vec{r}_{d',l'}^{\,0}-\vec{r}_{d',l'}^{\,g'}$
and $\vec{r}_j+\vec{r}_{d',l'}^{\,0}+\vec{r}_{d',l'}^{\,g'}$
are linked by a bond (and occupied by two independent $+1/2$ 
spinons) then a physical final spin configuration is
generated provided that the sites of real-space coordinates
$\vec{r}_j+\vec{r}_{d,l}^{\,0}-\vec{r}_{d,l}^{\,g}$ and
$\vec{r}_j+\vec{r}_{d,l}^{\,0}+\vec{r}_{d,l}^{\,g}$  
are (i) occupied by two independent $+1/2$ spinons 
or (ii) linked by a bond. In turn, when in the initial spin
configuration where these four-site two-bond operators 
act onto the sites of the spin effective lattice of real-space 
coordinates $\vec{r}_j+\vec{r}_{d',l'}^{\,0}-\vec{r}_{d',l'}^{\,g'}$
and $\vec{r}_j+\vec{r}_{d',l'}^{\,0}+\vec{r}_{d',l'}^{\,g'}$
are occupied by two independent $+1/2$ spinons (and  
linked by a bond) then a physical final spin configuration is
generated provided that their sites of real-space coordinates
$\vec{r}_j+\vec{r}_{d,l}^{\,0}-\vec{r}_{d,l}^{\,g}$ and
$\vec{r}_j+\vec{r}_{d,l}^{\,0}+\vec{r}_{d,l}^{\,g}$    
are linked by a bond (and occupied by two independent 
$+1/2$ spinons).
\vspace{0.25cm}

Such rules follow naturally from the definition of the subspace where 
the operators of Eqs. (\ref{g-s1+general})-(\ref{g-s-l}) act onto. The
main criterion is that such operators have been constructed to inherently 
generating a faithful representation provided that the corresponding
state (\ref{phi-+-s1}) represents the $N^h_{s1}=0$ configuration state 
and hence is identical to that given in Eq. (\ref{phi-+}) with the same 
value of $N_{s1}=N_{a_{s1}}^D$.
For instance, the second rule results from all partitions of the 
summation on the right-hand side of Eq. (\ref{phi-+})
a site of the spin effective lattice being linked to exactly only one 
site. Indeed, in a given partition no two-site one bonds
join the same site. Furthermore, the third rule refers to
four-site two-bond operators whose two two-site one-bond operators
belong to the same $s1$ bond-particle operator $g_{\vec{r}_{j},s1}^{\dag}$
or $g_{\vec{r}_{j},s1}$ yet correspond to different pairs of sites of the 
spin effective lattice. Such four-site two-bond operators appear in the
expressions of the $s1$ bond-particle operators
$[g_{\vec{r}_{j},s1}^{\dag}]^2$ or $[g_{\vec{r}_{j},s1}]^2$,
respectively, which as confirmed in 
Appendix A give zero when acting onto the
subspace which the operators of Eqs. (\ref{g-s1+general})-(\ref{g-s-l})
refer to. The point is that when in the initial spin
configuration where the operators 
$b_{\vec{r},s1,d,l,g}^{\dag}b_{\vec{r}',s1,d',l',g'}^{\dag}$ and
$b_{\vec{r},s1,d,l,g}b_{\vec{r}',s1,d',l',g'}^{\dag}$
(and $b_{\vec{r},s1,d,l,g}b_{\vec{r}',s1,d',l',g'}$ and
$b_{\vec{r},s1,d,l,g}^{\dag}b_{\vec{r}',s1,d',l',g'}$)
act onto the sites of the spin effective lattice of real-space 
coordinates $\vec{r}_j+\vec{r}_{d',l'}^{\,0}-\vec{r}_{d',l'}^{\,g'}$
and $\vec{r}_j+\vec{r}_{d',l'}^{\,0}+\vec{r}_{d',l'}^{\,g'}$
are occupied by two independent $+1/2$ spinons (and  
linked by a bond) the operator on the right-hand side
of $g_{\vec{r}_{j},s1}^{\dag}g_{\vec{r}_{j},s1}^{\dag}$ 
and $g_{\vec{r}_{j},s1}g_{\vec{r}_{j},s1}^{\dag}$
(and $g_{\vec{r}_{j},s1}g_{\vec{r}_{j},s1}$ and
$g_{\vec{r}_{j},s1}^{\dag}g_{\vec{r}_{j},s1}$) which such 
four-site two-bond operators belong to, respectively,
has transformed that configuration into an  intermediate virtual state
where such sites are linked by a bond 
(and occupied by two independent $+1/2$ spinons). 

The $N^h_{s1}=0$ configuration state is given
exactly by the same superposition of partitions 
when the real-space coordinates $\vec{r}_{j}$ 
of the $s1$ bond-particle operators
of Eq. (\ref{g-s1+general})
where $j=1,...,N_{a_{s1}}^D$ are chosen to refer to any
of the two sub-lattices of the spin effective lattice. 
That property is straightforwardly confirmed for the
$g$-primary partition represented in Fig. 3
and is fulfilled provided that one sums over
all possible partitions. Hence,
whether the above real-space coordinates are those of
one or the other of such sub-lattices, the occupancy
configuration associated with the creation
of the $N_{s1}= N_{a_{s}}^D/2$ bond particles 
is exactly the same. This confirms that the two 
corresponding choices for the $s1$ effective
lattice describe the same $N^h_{s1}=0$ configuration state so that 
there is a gauge structure. This holds in the limit $N_a\gg 1$
that the present description refers to. 

\subsection{Two-site one-bond weights}

The exact dependence of the absolute value 
$\vert h_{g}\vert$ of the coefficients 
appearing in Eqs. (\ref{phi-+}) and (\ref{g-s1+sum-rule}) 
on the link length $\xi_{g}$ remains an open problem. 
However, if one assumes that $\vert h_{g}\vert$ 
has the following simple power-law dependence on that length,
\begin{equation}
\vert h_{g}\vert =  {C\over \xi_{g}^{\alpha_{s1}}}
\, , 
\label{c-L-d}
\end{equation}
the link-length expression (\ref{xi-L}) and
normalization condition (\ref{g-s1+sum-rule}) alone
imply that $C^2$ be given by,
\begin{eqnarray}
C^2 & = & {a_s^{2\alpha_{s1}}\over 
4\sum_{N_d}
\sum_{N_{\bar{d}}}
[(1+2N_d)^2 + (2N_{\bar{d}})^2]^{-\alpha_{s1}}}
\,  ; \hspace{0.50cm} D=2 \, ,
\nonumber \\
C^2 & =  & {a_s^{2\alpha_{s1}}\over 
2\sum_{N_1}(1+2N_d)^{-2\alpha_{s1}}} \,  ; \hspace{0.50cm} D=1 \, .
\label{C-D-1-2}
\end{eqnarray}
Here the summations are over the range of $N_d$ and $N_{\bar{d}}$ 
(and $N_1$) values given in Eq. (\ref{link-sum}) for the square (and 1D) 
lattice. An expression of the general form (\ref{c-L-d})
would imply that the dependence of $\vert h_{g}\vert$
on $U/4t$ and $x$ occurs through that of the
exponent $\alpha_{s1} =  \alpha_{s1} (U/4t,x)$
only. 
\begin{figure}
\includegraphics[width=7cm,height=3.5cm]{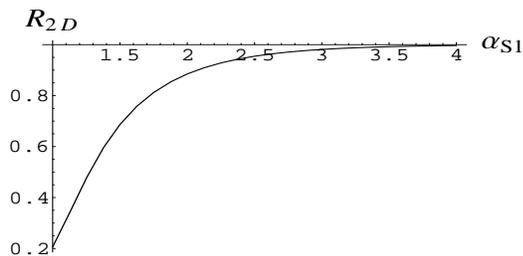}
\caption{The relative spectral weight of the primary links
$R_{2D}$ of Eq. (\ref{rela-weight}) for the
square lattice as a function of the exponent $\alpha_{s1}$
for the range $\alpha_{s1}\geq 1$. That
relative weight refers to
the simple power-law expressions given in 
Eqs. (\ref{c-L-d}) and (\ref{C-D-1-2}) for the coefficients
$\vert h_{g}\vert$ appearing in the 
summations of Eq. (\ref{g-s1+general}).
The physical range of the exponent $\alpha_{s1}$ corresponds
typically to $R_{2D}>0.9$ so that the primary
links are behind most of the $s1$ bond-particle
spectral weight. For $\alpha_{s1}\approx 2.5$ the
ratio $R_{2D}$ is larger than $0.9$ and approaches
quickly the unit upon further increasing $\alpha_{s1}$.}
\end{figure}
 
For the coefficients $\vert h_{g}\vert$ power-law expressions given in 
Eqs. (\ref{c-L-d}) and (\ref{C-D-1-2}) the ratio of the spectral weight 
of a primary link over the weight of the
corresponding link family,
\begin{equation}
R_{z} \equiv {\vert h_0\vert^2\over
\sum_{g=0}^{[N_{s1}/2D-1]} \vert h_{g}\vert^2} 
= 2D\,\vert h_0\vert^2
\,  , \hspace{0.50cm} z=1D,2D \, ,
\label{ratios-2-1}
\end{equation}
is given by,
\begin{eqnarray}
R_{2D} & = & {1\over \sum_{N_d}\sum_{N_{\bar{d}}}
[(1+2N_d)^2 + (2N_{\bar{d}})^2]^{-\alpha_{s1}}}
\,  ; \hspace{0.15cm} D=2 \, ,
\nonumber \\
R_{1D} & = & 
{1\over\sum_{N_1}(1+2N_d)^{-2\alpha_{s1}}}\,  ; \hspace{0.15cm} D=1 \, .
\label{rela-weight}
\end{eqnarray}
The ratios of Eq. (\ref{rela-weight}) refer to the limit 
$N_a\rightarrow\infty$ and are plotted in Figs. 5 and 6
for the square and 1D lattice, respectively, as a function
of the exponent $\alpha_{s1}$ for the range $\alpha_{s1}\geq 1$.
For  $\alpha_{s1}$ slightly larger than
$5D/4$ where $D=1,2$ the ratios $R_{1D}$ ($D=1$) and $R_{2D}$ 
($D=2$) are larger than $0.9$. 

For $N_a\rightarrow\infty$ the exact coefficients $\vert h_{g}\vert$
are decreasing functions of the link length whose expressions most likely 
are not, at least for the whole link-length range, of the simple form given in Eqs. (\ref{c-L-d}) and 
(\ref{C-D-1-2}). However the exact sum-rules (\ref{g-s1+sum-rule}) 
together with the coefficients $\vert h_{g}\vert$ being decreasing
functions of the link length implies that at least for link-length not too small the 
exact coefficients $\vert h_{g}\vert$ fall off as in Eq. (\ref{c-L-d}) and
the corresponding ratios of Eq. (\ref{ratios-2-1}) have a behaviour similar to that
shown in Figs. 5 and 6. Hence for the square lattice one expects that 
$\alpha_{s1}>2.5$ so that the ratio $R_{2D}$ is larger than $0.9$ 
and the primary links are behind most of the spectral weight of the
$s1$ bond particle, alike for the 1D lattice for $\alpha_{s1}>1.25$.

Here we considered the $N^h_{s1}=0$ configuration state 
where all sites of the $s1$ effective lattice are occupied. 
In the following we generalize our results to configuration
states with a finite number of unoccupied sites. 
\begin{figure}
\includegraphics[width=7cm,height=3.5cm]{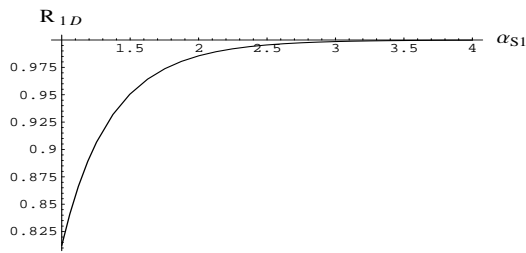}
\caption{The relative spectral weight of the primary links
$R_{1D}$ of Eq. (\ref{rela-weight}) for the
1D lattice as a function of the exponent $\alpha_{s1}$
for the range $\alpha_{s1}\geq 1$. 
The ratio $R_{1D}$ approaches the unit upon increasing
the value of $\alpha_{s1}$ more quickly than
the ratio plotted in Fig. 5 for the square lattice.
For $\alpha_{s1}=n_{s1}=1,2,3,...$ integer 
the relative weight reads $R_{1D}=1/[(1-2^{-2n_{s1}})\zeta (2n_{s1})]$
where $\zeta (x)$ is a Riemann's zeta function. For
instance $R_{1D}=8/\pi^2\approx 0.811$ for 
$\alpha_{s1}=1$ and $R_{1D}=96/\pi^4\approx 0.986$ 
for $\alpha_{s1}=2$.}
\end{figure}

\section{The $N^h_{s1}=1,2$ configuration states and the
kink-like and anti-kink-like link occupancies associated with 
the unoccupied sites}

The one- and two-electron subspace considered 
in this paper is spanned by states with none, one, and two $s1$ effective-lattice
unoccupied sites. There are several configuration states
with one or two unoccupied sites in that lattice so that
such sites can move around in it.
We recall that such a motion is independent of that of
the rotated electrons that singly occupy sites 
of the original lattice relative
to the rotated-electron unoccupied and doubly-occupied  
sites of that lattice. The latter are instead described by the motion
of the $c$ fermions whose occupancy configurations
correspond to the state representations of the charge
global $U(1)$ symmetry. In turn, the configuration states
associated with the motion around in the $s1$ and
effective lattices of the $s1$ bond particles
relative to their unoccupied sites refer to state
representations of the global spin $SU(2)$ symmetry.
The global $\eta$-spin $SU(2)$ symmetry does not play
any role in the one- and two-electron subspace, since
it corresponds to a single occupancy configuration ($x>0$) 
or none occupancy configurations ($x=0$) for
all states which span that subspace. For $x>0$ this is
the $\eta$-spin vacuum on the right-hand side 
of Eq. (\ref{vacuum}), which is invariant under the
electron - rotated-electron unitary transformation.
For $x=0$ one has that $N_{a_{\eta}}^D=0$ so that 
the $\eta$-spin effective lattice does not exist.

Analysis of the transformation laws of the objects whose occupancy configurations generate
the energy eigenstates of the square-lattice model in the one- and two-electron subspace 
\cite{companion} reveals that the excited states
generated by application onto $x\geq 0$ and $m=0$ ground states 
of one- and two-electron operators have one unoccupied site
and none or two unoccupied sites in the $s1$ effective
lattice, respectively. The energy eigenstates that span 
such a subspace are generated in Ref.
\cite{companion} by $c$ and $s1$ fermions
momentum occupancy configurations.
The $s1$ fermion occupancies in the $s1$
momentum band introduced in Ref. \cite{companion}
that generate energy eigenstates with one and two $s1$ band holes  
can be expressed as suitable superpositions of the set 
of $N_{s1}^h=1$ and $N_{s1}^h=2$ configuration states, 
respectively, studied in the following. 

The structure of the $N^h_{s1}=0$ configuration states is simpler than
that of the $N^h_{s1}=1,2$ configuration states. A key
property simplifies the study of the latter states:
There is an one-to-one correspondence between the
occupancies of the $N_{s1}$ occupied sites of the $s1$ effective lattice 
of the partitions of a $N^h_{s1}=0$ configuration state 
and those of the occupied sites of the $s1$ effective lattice
of partitions of the $N^h_{s1}=1,2$ configuration states
with the same number $N_{s1}$ of $s1$ bond particles. The point
is that except for a change in the real-space coordinates 
of some of the sites, the link configurations of the sites
of the spin effective lattice in one-to-one correspondence
with each other remain unaltered. 

\subsection{$N^h_{s1}=1$ $g$-primary partitions
of the four sub-configuration states}

We start our analysis by considering
the configuration states with one unoccupied site.
The $s1$ fermion momentum occupancy of
excited $[N_{\uparrow}-N_{\downarrow}]=1$ states 
of $x\geq 0$ and $m=0$ ground states generated by 
application of one-electron operators onto the 
latter states is described by a suitable superposition of the
set of $N^h_{s1}=1$ configuration states studied in the
following. For such configuration states the unoccupied site 
is associated with their single independent $+1/2$ spinon. 
For simplicity we have considered above that 
for the $N^h_{s1}=0$ configuration state the
number of sites $N_{a_s}^D=2S_c=N_c$ of the
spin effective lattice is such that $N_{a_s}$
is for $D=2$ an integer so that
the spin effective lattice is a square lattice.
In turn, for configuration states with one unoccupied site 
the number $N_{a_s}^D$ of sites of the spin effective lattice
is odd so that for $D=2$ the designation $N_{a_s}^D$ is
not to be understood to imply that the number $N_{a_{s}}$ 
is an integer. 

We recall that the number of sites of the spin effective lattice
is given by $N_{a_s}^D=2S_c=N_c$ where $S_c$ is the eigenvalue
of the generator of the global $U(1)$ symmetry 
${\tilde{S}}_c= {\hat{V}}^{\dag}\,{\hat{S}}_c\,{\hat{V}}$,
${\hat{S}}_c= {\hat{Q}}/2$, the operator
${\hat{Q}}$ is given in Eq. (\ref{Q-op}), and
$N_c$ is the number of $c$ fermions. Since $S_c$ is
a half-off-integer, according to the studies of Ref. \cite{bipartite} 
the global $SO(3)\times SO(3)\times U(1)$ symmetry of the present quantum
problem implies that the $\eta$-spin $S_{\eta}$ and
spin $S_s$ are half-off-integers as well. It follows that the number of 
sites of the $s1$ effective lattice is according to Eq. (\ref{Nas1-Nhs1})
given by $N_{a_{s1}}^D = [N_{a_s}^D/2 + S_s]$ where $S_s=1/2$.
Out of those, $N_{s1}=[N_{a_{s1}}^D  -S_s]$ are occupied by
$s1$ bond particles and one is unoccupied, consistently with $N^h_{s1}=2S_s=1$.
Hence such states have also one unoccupied site in the $s1$ 
effective lattice. 

For the study of the $N^h_{s1}=0$ configuration state
we considered in Section 3 a change of gauge structure  
such that the real-space coordinates of the sites of the 
$s1$ effective lattice correspond to one of the two
sub-lattices of the spin effective lattice. The results of Ref.
\cite{companion} confirm that for the
limit $N_a^D\gg 1$ and hole concentrations 
$x$ such that $(1-x)\geq 1/N_a^D$ the two choices of $s1$ effective
lattice lead to the same quantum numbers for the
whole one- and two-electron subspace.

Let us consider $N^h_{s1}=1$ configuration states
whose spin effective lattice has $2N_{s1}+1$
sites and for the model on the square lattice the spin effective 
lattice of the corresponding $N^h_{s1}=0$ 
configuration state with the same
number $N_{s1}$ of $s1$ bond particles is a square lattice 
and has $N_{a_{s}}\times N_{a_{s}}=2N_{s1}$ sites.
For the present $N_{a_{s}}^D\rightarrow\infty$
limit the correct physics is then 
achieved if one considers that out of the
$2N_{s1}+1$ sites of the spin effective
lattice of such $N^h_{s1}=1$ configuration states,
$2N_{s1}$ sites correspond to a square lattice
and the position of the extra site is well defined
and given below. The latter site does not belong to the square
lattice formed by the remaining $2N_{s1}$ sites
of the spin effective lattice and has 
suitable boundary conditions such that it
is the unoccupied site of the $s1$ effective
lattice. However, for the spin effective lattice the
extra site and the unoccupied site are not
always the same site, as discussed below.

For the square (and 1D) lattice the extra site boundary
conditions are compatible and consistent
with the periodic boundary conditions of each row
and column (and chain). For the $N^h_{s1}=0$ 
configuration state of the model on the square lattice 
such a row and column periodic boundary 
conditions imply torus periodic boundary conditions
for the spin effective lattice. Here the presence of
the extra site slightly affects the latter boundary
conditions. Hence the spin effective lattice of
$N^h_{s1}=1$ configuration states has no
{\it pure} torus periodic boundary conditions.
Within the extra-site boundary conditions given below
each $N^h_{s1}=1$ configuration state has for 
the square (and 1D) lattice four (and two) 
{\it sub-configuration states}. The spin effective 
lattice is the same for all such sub-configuration states. 
The concepts of a partition and $g$-partition remain the same as
for the corresponding $N^h_{s1}=0$ configuration state,
except that now there is an extra site. 

For the model on the square lattice
the extra site of the spin effective lattice belongs 
to the row or column that for a $d=1$
and $d=2$ primary partition specified below,
respectively, the unoccupied site 
belongs to. The real-space coordinate of the extra site and its
Cartesian coordinates are denoted by,
\begin{equation}
{\vec{r}}_{j_{0}} = [x^{0}_1,x^{0}_2] \, .
\label{unocc}
\end{equation}  
In the present $N_a\rightarrow\infty$ 
limit the following periodic conditions yield the 
correct physics and assure that the extra site 
is uniquely defined and the same site for all partitions
of a given $N^h_{s1}=1$ configuration state,
\begin{equation}
x^{0}_d = x^{0}_d + l\,(L + a) \, ;
\hspace{0.50cm} x^{0}_{\bar{d}} = 
x^{0}_{\bar{d}} \, ; \hspace{0.50cm} d=1,2 
\, , \hspace{0.15cm} l=\pm 1 \, ,
\label{x-0-d}
\end{equation}
where the index $\bar{d}=1,2$ is defined as in Eq.
(\ref{xd-xd}). Such boundary conditions apply to the 
square lattice for $d=1,2$ and $l=\pm 1$ and to the 1D 
lattice for $d=1$ and $l=\pm 1$. For
the row $(d=1)$ [and column $(d=2)$ for the
model on the square lattice] the extra site belongs to, they 
are equivalent to periodic boundary 
conditions. Indeed that row [and column] has
$N_a+1$ sites and hence its length
is $L + a=(N_a+1)\,a$. 

For each sub-configuration state there is exactly
one $g$-primary partition, which is a superposition of
the corresponding four primary partitions.
Hence instead of $2D=2,4$ primary partitions, for a $N^h_{s1}=1$ configuration
state there are $8D=8,16$ primary partitions. 
For all partitions of the summation on the
right-hand side of Eq. (\ref{phi-+}) the extra site has
the same Cartesian coordinates. Nonetheless, the
extra site and the unoccupied site are not
the same site of the spin effective lattice for all such partitions.
For instance, the extra site and the unoccupied site are
the same site for three out of the four primary partitions of a primary $g$-basic 
partition, as confirmed below.

For the model on the square lattice the periodic boundary conditions of 
the sites belonging to rows and columns other than that 
of the extra site refer to a length $L$ rather than 
$L+a$. For the extra site the boundary conditions (\ref{x-0-d})
apply to both the square and 1D lattice and
are valid for the $2D=2,4$ sub-configuration states 
considered below. In turn, for the model on the square lattice
and concerning the remaining sites of the 
extra-site row and column such periodic boundary 
conditions apply to those belonging to the row for horizontal
sub-configuration states and to the column for vertical
sub-configuration states, respectively. For horizontal
(and vertical) sub-configuration states the sites of the 
extra-site column (and row) obey periodic boundary
conditions associated instead with the length $L$, alike the
sites that do not belong to the extra-site column 
and row. That the periodic boundary conditions of
one row or column of the spin effective lattice
refer to the length $L+a$ rather than $L$ is 
for the model on the square lattice behind the above deviation
from the pure torus periodic boundary conditions.

Except for the unoccupied site, the remaining
$2N_{s1}$ sites of the spin effective lattice 
are occupied by $N_{s1}$ $s1$ bond particles. 
Furthermore, out of the $2N_{s1}+1$ sites of that 
lattice, for the $2N_{s1}$ sites 
other than the extra site, which for some partitions is not the unoccupied
site, there are two well-defined sub-lattices, alike for the
$N^h_{s1}=0$ configuration state. Indeed, the extra site
belongs to the spin effective lattice but does not belong to
any of these two sub-lattices. Moreover, now sub-lattice 1
is that whose real-space coordinates are the same as those
of the sites of the $s1$ effective lattice whose 
row or column are not that of the extra
site. In turn, for horizontal (and vertical) sub-configuration
states the occupied sites of the $s1$ effective lattice 
of the extra-site row (and column) belong instead 
to sub-lattice 2. 

The $s1$ bond-particle operators act onto a subspace
with constant values $N_{a_s}^D=2S_c=N_c$ of spin-effective-lattice 
sites. Therefore, suitable analysis of the general
expressions of the number of sites and
unoccupied sites of the $\alpha\nu$ effective
lattices \cite{companion}
reveal that for $\alpha\nu =s1$ one of these operators
or products of such operators can generate transitions
where the value of $N_{s1}^h$ either is conserved
or changes by an even number. Indeed, transitions
where such a value changes by an odd number
imply that that of the number $N_{a_s}^D=2S_c=N_c$ of both 
spin-effective-lattice sites and $c$ fermions also changes.
For instance, transitions from the $N^h_{s1}=0$ configuration 
state to a $N^h_{s1}=1$ configuration state involve
creation or annihilation of one electron. Hence they
involve creation or annihilation of one rotated 
electron as well. In terms of
the objects of our description that process involves  
creation or annihilation, respectively, of a $c$ fermion
plus creation of an independent spinon  \cite{companion}. In the
LWS representation the latter is an
independent $+1/2$ spinon whose creation is
equivalent to adding an extra site to the spin effective lattice.

Therefore, addition of the extra site is a process
beyond the $s1$ bond-particle algebra whose
operators are well defined for subspaces
with constant values of $N_{a_s}^D=2S_c=N_c$ spin-effective-lattice sites only. 
Fortunately, our general description also
accounts for such a process, which involves both
creation or annihilation of a $c$ fermion and
creation of an independent $+1/2$ spinon.
In turn, the generators of the transitions between 
different $N^h_{s1}=1$ configuration states involve
$s1$ bond-particle operators only.
Such transitions are equivalent to the motion
around in the spin effective lattice of $s1$
bond particles. But since in the present case
there is a single unoccupied site it is often
more convenient to describe such a motion
in terms of those of that site.
 
The unoccupied site has the same position
in the $s1$ effective lattice and $g$-primary partitions
of the $2D=2,4$ sub-configuration states. Such a position
is the same as that of the extra
site in the spin effective lattice. For each position 
of the extra site in that lattice there is a $N^h_{s1}=1$ 
configuration state. Specifically, there are $N_{a_{s1}}^D = 
[N_{a_s}^D/2 + S_s] =N_{s1}+1$ such
configuration states. The $s1$ fermion occupancies
of Ref. \cite{companion} of the above mentioned excited states generated 
by application onto $x\geq 0$ and $m=0$ ground states of one-electron operators can be
expresed as suitable superpositions
of such $N_{a_{s1}}^D = 
[N_{a_s}^D/2 + S_s] =N_{s1}+1$ configuration states.
For the model on the square lattice the four 
sub-configuration states of each of such $N^h_{s1}=1$ 
configuration states are called {\it horizontal-kink}, {\it horizontal-anti-kink}, 
{\it vertical-kink}, and {\it vertical-anti-kink sub-configuration state}. 

It useful for the description of the $N^h_{s1}=2$ 
configuration states of Subsection 4-5 to introduce 
the $N^h_{s1}=1$ $g$-primary 
partition of the horizontal-kink sub-configuration state 
as a limiting case of a general configuration whose
unoccupied site and extra site do not have the same real-space
coordinate in the $s1$ effective lattice. 
The $g$-primary partition  of the 
above sub-configuration state is then reached by considering 
the particular case when the unoccupied and extra site
are the same site. As discussed below,
the unoccupied site of a $g$-primary partition (and that of the underlying $s1$ effective lattice) 
is identified with the unoccupied site of three out of its  
four primary partitions. A sub-domain 
of the spin-effective lattice with a part of such a general 
$g$-primary partition  is shown in Fig. 7 for 
the square lattice. Its extra site is not
shown in the sub-domain under consideration and is not the unoccupied 
site marked by the open circle. The occupied sites of the $s1$ 
effective lattice are represented by filled circles. 
\begin{figure}
\includegraphics[width=7cm,height=5cm]{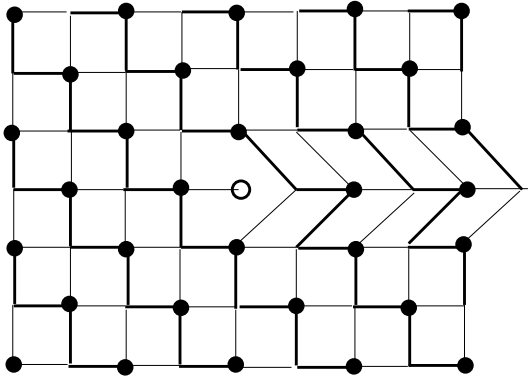}
\caption{Sub-domain of the spin-effective lattice representing 
a general type of link occupancy for the model on the square lattice which when the extra and unoccupied
sites have the same real-space coordinate corresponds to the
$g$-primary partition  of the 
$N^h_{s1}=1$ horizontal-kink sub-configuration state.
However, in the link configuration shown here
the unoccupied site does not coincide with the
extra site so that the former site plays the role
of a mobile domain wall, alike for some
$g$-primary partitions of
$N^h_{s1}=2$ configuration states considered in Subsection 4-5.
The horizontal (and vertical) thin and thick lines refer 
to $d=1$ (and $d=2$) $l=+1$ and $l=-1$ primary links, 
respectively.  For the $d=1$, $l=+1$ partition the unoccupied site is that attached to three 
links. In turn, for the $d=2$, $l=\pm 1$ partitions and $d=1$, $l=-1$ partition the unoccupied site is that which has
a single link attached to it. The latter site is marked by
an open circle in the figure and plays the role of unoccupied 
site of the $s1$ effective lattice. The occupied sites of 
that lattice are represented by filled circles. The
extra site does not belong to the sub-domain
shown here and marks the end of the horizontal kink-like
link occupancy located on the right-hand side
of the unoccupied site.}
\end{figure}

The row kink-like link occupancy starting on the right-hand side of 
the unoccupied site ends at the extra site. We say that such a link 
occupancy and those of the surrounding rows have opposite signs. 
Indeed, the $s1$ bond-particle sites of the row part associated
with the former occupancy belong to sublattice 2 of the spin 
effective lattice whereas those of the $s1$ bond particles 
on the surrounding rows belong to sublattice 1.

The $g$-primary partition  of the 
horizontal-kink sub-configuration state 
corresponds to a particular case of that represented in Fig. 7
whose extra and unoccupied site are the same site so that the 
kink-like link occupancy extends over the whole unoccupied-site 
row. Figure 8 shows a sub-domain of the spin-effective lattice representing 
the $g$-primary partition  of the 
$N^h_{s1}=1$ horizontal-kink sub-configuration state
for the square lattice. Since one of the issues 
to be further discussed is the periodic boundary conditions of
the extra-site row, for simplicity the $N_{a_s}\gg 1$ 
sites of a row are represented in Fig. 8 by a few sites.  
The last site of each row is marked by a $X$
to indicate that it is the same site as the first
site of the same row. In turn, for the vertical direction the figure
represents a sub-domain only so that no 
equivalent sites are included, alike for the sub-domain
shown in Fig. 7. Note that for the extra-site row 
of Fig. 8 the extra site is marked by a $X$
and thus is the same as the unoccupied
site marked by an open circle. Therefore and
in contrast to Fig. 7, in Fig. 8 the kink-like link configuration 
extends over the whole extra-site row. 

Both for the $d=2$, $l=\pm1$ primary partitions and 
$d=1$, $l=-1$ primary partition of the 
$g$-primary partition shown in Fig. 8 and that of Fig. 7 the 
unoccupied site is the site attached to a single link, 
marked by the open circle in both figures. The $d=1$, $l=+1$ 
primary partition is the only one for which the
unoccupied site is that attached to three links in
the figures. The site marked by the open circle plays
the role of the unoccupied site of both the 
$g$-primary partition and $s1$ effective
lattice of the four sub-configuration states. 

For the $g$-primary partition  
of the horizontal-kink sub-configuration state we call 
{\it kink row} or {\it $1,+1$ line} the link occupancy of
the sites belonging to the same row as the unoccupied site. 
The notation $d,l$ line where $d=1$ and
$l=+1$ stems from the $d=1$, $l=+1$ primary 
partition  being the only one out of the
four such configurations of the corresponding
$g$-primary partition for which 
the unoccupied site is attached to three 
links. For the $g$-primary partition  
of each of the four sub-configuration states of
a $N_{s1}^h=1$ configuration state there is
a $d,l$ line and in each case the corresponding
$d,l$ primary partition  
is that for which the unoccupied site is attached 
to three links. Furthermore, for the remaining
three primary partitions the unoccupied 
site is always that attached to a single link:
It is the extra site, as illustrated in Fig. 8. 

We call {\it $d,l$ $g$-primary partition} that which involves a $d,l$ line. 
Let ${\vec{r}}_{j_{un}}$ denote the real-space coordinate of the 
unoccupied site in the spin 
effective lattice for each of the four primary partitions of 
the $d,l$ $g$-primary partition. For each of such primary 
partitions it is given in terms of the real-space coordinate 
${\vec{r}}_{j_{0}}$ of the extra site provided in Eq.
(\ref{unocc}) as follows,
\begin{eqnarray}
{\vec{r}}_{j_{un}} & = & {\vec{r}}_{j_{0}} \, ; \hspace{0.15cm}
\bar{d},\pm l \hspace{0.10cm} {\rm and} \hspace{0.10cm}
d,-l \hspace{0.10cm} {\rm primary} \hspace{0.10cm}
{\rm partitions} \, ,
\nonumber \\
{\vec{r}}_{j_{un}} & = & {\vec{r}}_{j_{0}} + 2{\vec{r}}_{d,l}^{\,0} \, ; \hspace{0.15cm}
d,l \hspace{0.10cm} {\rm primary} \hspace{0.10cm}
{\rm partition} \, ,
\label{r-jun}
\end{eqnarray}
where the index $\bar{d}$ is defined as in Eq. (\ref{xd-xd}).  
\begin{figure}
\includegraphics[width=4.5cm,height=5cm]{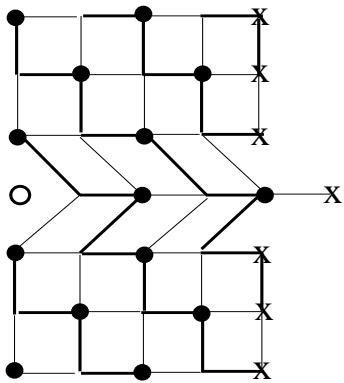}
\caption{Sub-domain of the spin-effective lattice representing 
the $g$-primary partition  of the 
$N^h_{s1}=1$ horizontal-kink sub-configuration state
for the model on the square lattice. The horizontal (and vertical) thin and thick lines refer 
to $d=1$ (and $d=2$) $l=+1$ and $l=-1$ primary links, 
respectively, the unoccupied site is marked by
an open circle, and the sites corresponding to
real-space coordinates of $s1$ bond particles by  
filled circles. For the vertical direction
only part of the configuration is shown. In turn, for the
horizontal direction the $N_{a_s}\gg 1$ sites of a row
are represented by a few sites. Indeed, one 
of the goals of the figure is to illustrate the row periodic 
boundary conditions, a site marked by a $X$
being the same as the first site on the left-hand
site of the same row. For instance, the unoccupied
site marked by an open circle is the same site as 
the last site on the right-hand site of the same
row. Note that the extra-site
row has one more site than the remaining rows
and such a site is both the extra site and
the unoccupied site so that the
kink-like link configuration extends over the whole
extra-site row, in contrast to the configuration shown 
in Fig. 7. On the other hand, the link configurations 
associated with the unoccupied site are the same
as in Fig. 7. For the $d=1$, $l=+1$ primary partition the 
unoccupied site is that attached to three 
links and for the $d=2$, $l=\pm 1$ primary partitions and 
$d=1$, $l=-1$ primary partition that which has a single link 
attached to it.}
\end{figure}

Figure 9 shows four small sub-domains of the spin-effective 
lattice representing part of general $g$-primary partitions of the 
type represented in Fig. 7. As discussed
below in Subsection 4-5, such sub-domains involve unoccupied-site mobile
domains walls which occur in $N^h_{s1}=2$ sub-configuration
states when the two unoccupied sites belong to the
same row or column. However, the link structure
associated with the unoccupied site at the center of each of
such sub-domains is that also occurring in the $g$-primary partitions 
of the four $N^h_{s1}=1$ sub-configuration states, respectively.
In turn, each of these sub-domains includes four primary links of a $s1$
bond particle with a single link attached to the unoccupied site. 
As confirmed for the horizontal-kink sub-configuration state on 
comparing it with the link occupancy configurations of Fig. 8, the 
only difference relative to the $N^h_{s1}=1$ sub-configuration states
refers to the structure and position of the remaining 
three primary links of that $s1$ bond particle, the link-structure of 
the unoccupied site being the same as that shown in Fig. 9.  
\begin{figure}
\includegraphics[width=3.5cm,height=2.65cm]{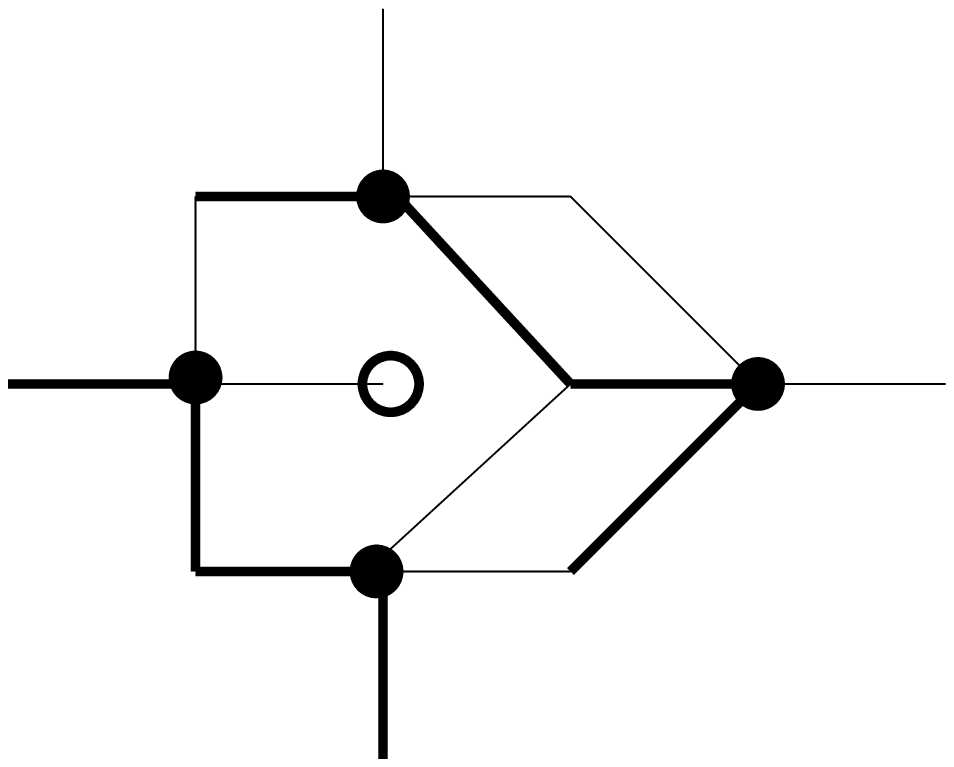}
\hspace{0.2cm}
\includegraphics[width=3.5cm,height=2.65cm]{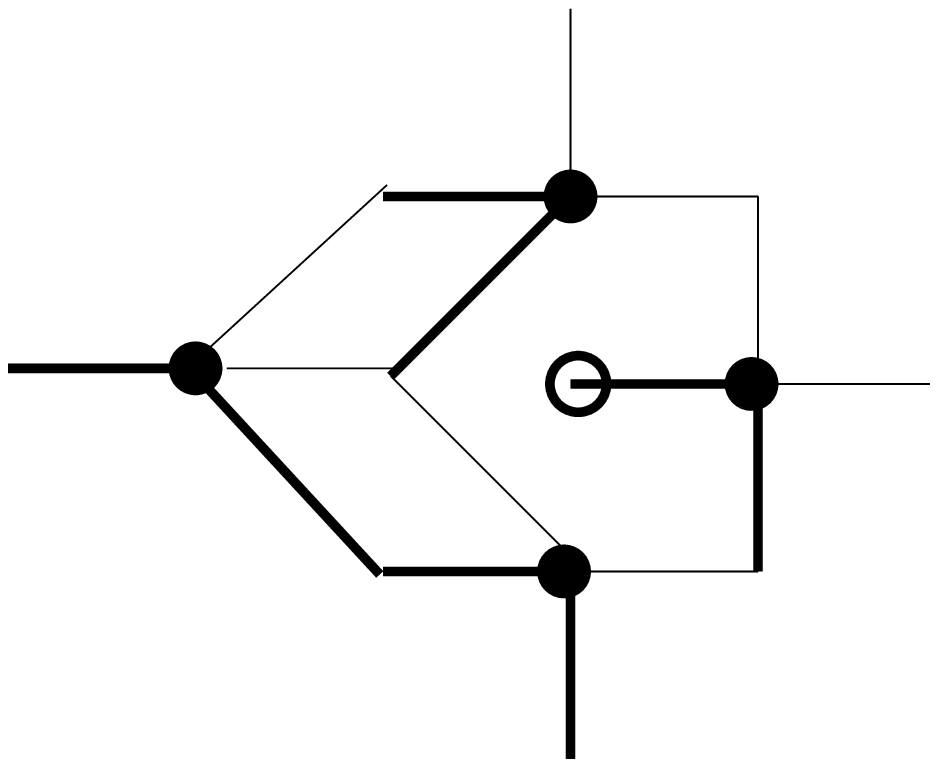}
\includegraphics[width=2.65cm,height=3.5cm]{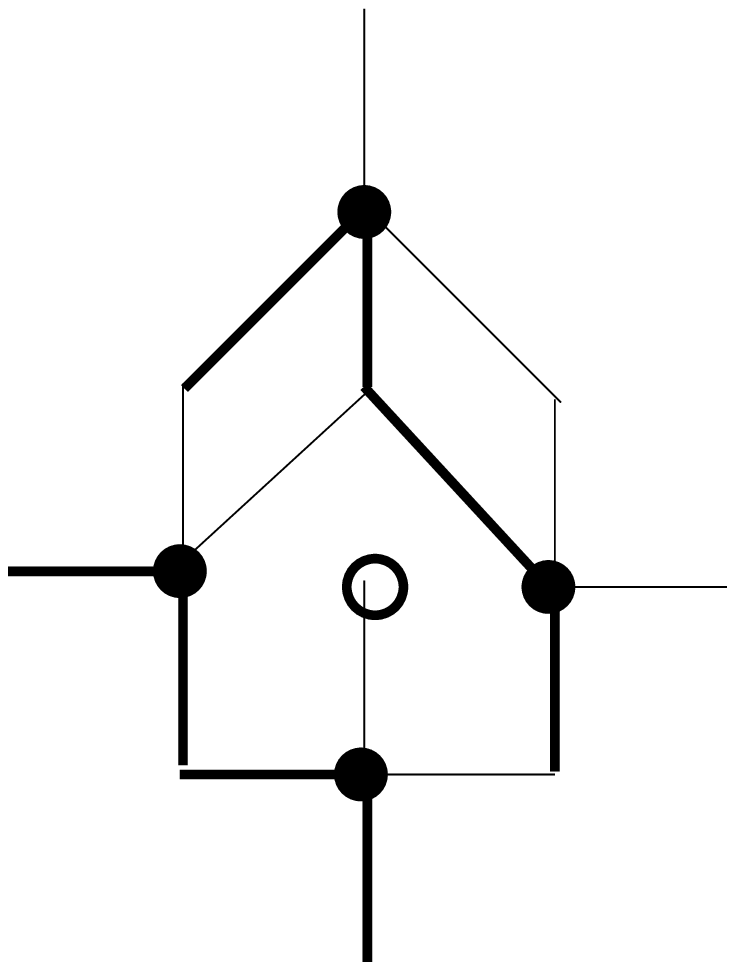}
\hspace{0.2cm}
\includegraphics[width=2.65cm,height=3.5cm]{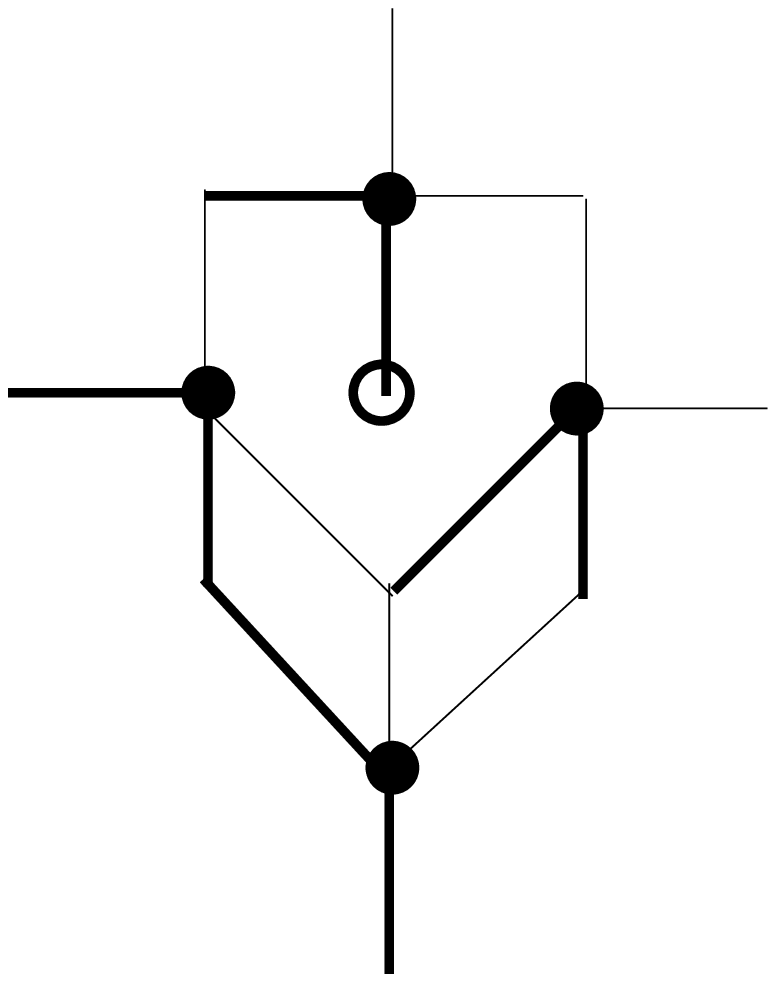}
\caption{Four small sub-domains of the spin-effective lattice 
showing part of $g$-primary partitions.
Alike in Fig. 7, each sub-domain includes an unoccupied site which plays 
the role of a mobile domain wall. The first two configurations of the figure refer to 
kink-like and anti-kink-like link configurations, respectively,
of horizontal sub-configuration states. The two last configurations correspond to 
kink-like and anti-kink-like link configurations, respectively,
of vertical sub-configuration states. Such mobile domain 
walls occur in some $N^h_{s1}=2$ sub-configuration states
with two unoccupied sites on the same row or column rather than
in $N^h_{s1}=1$ sub-configuration states. However, the
link structure of the unoccupied site is the same as
for the latter states.}
\end{figure}

Specifically, the unoccupied-site link structure of the top 
configurations on the left- and right-hand side of 
Fig. 9 are the same as those of the 
$g$-primary partitions of
$N^h_{s1}=1$ horizontal-kink
and horizontal-anti-kink sub-configuration states, respectively.
For the $d=2$, $l=\pm1$ primary partitions 
and $d=1$, $l=-1$ (and $d=1$, $l=+1$) primary partition on the left-hand side  (and right-hand site) 
of the figure top 
the unoccupied site is that attached to a single link, marked 
by an open circle. In turn, for the $d=1$, $l=+1$ 
(and $d=1$, $l=-1$) primary partition  
the unoccupied site is that attached to three links.
Moreover, for the $d=1$, $l=\pm1$ primary partitions 
and the $d=2$, $l=-1$ (and $d=2$, $l=+1$) primary partition on the left-hand side 
(and right-hand side) of the figure lower-limit 
associated with the vertical-kink (and vertical-anti-kink) sub-configuration 
state the unoccupied site is that attached to a single link, marked 
by an open circle. On the other hand, for the corresponding $d=2$ and 
$l=+1$ (and $d=2$, $l=-1$) primary partition  
the unoccupied site is that attached to three links. Such four
types of unoccupied-site link structures correspond to the
$g$-primary partitions of
the four $N^h_{s1}=1$ sub-configuration states,
respectively. 

The $g$-primary partitions
of the horizontal-kink and horizontal-anti-kink (and vertical-kink and
vertical-anti-kink) sub-configuration states include
a kink row as that represented in Fig. 8
and a {\it anti-kink row} (and a {\it kink column} and
{\it anti-kink column}), respectively. Such a row (and
column) link occupancy configuration involves the set of sites 
belonging to the same row (and column) as the 
extra site. As mentioned above, an alternative
designation for the kink row, anti-kink row, kink column,
and anti-kink column is $1,+1$ line,
$1,-1$ line, $2,+1$ line, and $2,-1$ line, respectively.
The general $d,l$ line occurs in a primary $g$-basic 
partition for whose $d,l$ primary partition  
the unoccupied site is attached to three links.

There is an apparent contradiction concerning the
following issue. On the one hand, that
the extra-site periodic boundary conditions (\ref{x-0-d})
refer to both a row and column of $N_{a_s}+1$ sites 
is valid for the $2D=2,4$ sub-configuration states simultaneously.
On the other hand, concerning the $g$-primary partitions 
of such states, $N_{a_s}+1$-site periodic boundary 
conditions apply to either the sites of the spin effective lattice other 
than the extra site belonging to the extra-site 
row or extra-site column, respectively. Specifically, for 
$g$-primary partitions of horizontal (and vertical) sub-configuration
states such periodic boundary conditions 
apply to the sites belonging to the
$1,l$ (and $2,l$) line only, whereas for the sites belonging
to the extra-site column (and row) other than the extra site
the periodic boundary conditions refer to $N_{a_s}$ sites 
rather than to $N_{a_s}+1$ sites. 

The reason behind that
apparent inconsistency is that the link occupancy configurations
of each of the $2D=2,4$ $d,l$ lines of the   
$2D=2,4$ $g$-primary partitions, respectively,
are different and thus refer to different sub-configuration
states yet they all involve the same extra site. We recall
that the extra site of Figs. 7-9 is linked to one site only. 
For $g$-primary partitions with a
$1,l$ (and $2,l$) line the extra site is linked to
a site belonging to that line only so that only the
real-space coordinates of the remaining sites belonging to 
the $1,l$ (and $2,l$) line are shifted, whereas those
of the sites belonging to the extra-site column (and
row) remain unaltered. 

In addition to the $d,l$ line of a $g$-primary partition, it is
useful to consider its two nearest neighboring
lines. For each such an occupancy configuration
there are three $d,l,l'$ lines where
$l'=-1,0,+1$. Here $d,l,0\equiv d,l$ line is a
more general notation for the $d,l$ line.
For a kink or anti-kink row the
$1,l,l'$ line is the row above it for $l'=+1$ 
and below it for $l'=-1$. Moreover,
for a kink or anti-kink column the
$2,l,l'$ line is the column on its right-hand side 
for $l'=+1$ and on its left-hand side for $l'=-1$. 
 
\subsection{Motion of the unoccupied site around in the 
spin effective lattice}
 
The motion of the unoccupied site around in the 
spin effective lattice discussed here can be described
as well in terms of the motion of the corresponding 
unoccupied site of the $s1$ effective lattice around in
such a lattice. In the case of the $s1$ fermion description
of the square-lattice model investigated in Ref. \cite{companion} the generators
of the motion considered here are the two $s1$ translation generators 
in the presence of the fictitious magnetic field ${\vec{B}}_{s1}$  
associated with the Jordan-Wigner transformation that maps
the $s1$ bond particles into $s1$ fermions. The components ${q_j}_{x1}$ and 
${q_j}_{x2}$ of the discrete momentum values ${\vec{q}}_j$ of the $s1$ fermions are the 
eigenvalues of such two $s1$ translation generators. Since the $s1$ fermion 
creation and annihilation operators 
are defined in and act onto subspaces spanned by mutually neutral states \cite{companion}
such operators commute and the two components ${q_j}_{x1}$ and 
${q_j}_{x2}$ can be simultaneously specified.
  
It follows from the above analysis that for the $s1$
effective lattice and the $2D=2,4$  
$g$-primary partitions of all
sub-configuration states the unoccupied site
has a well-defined position. 
For $N^h_{s1}=1$ there are $N_{a_{s1}}=N_{s1}+1$ 
configuration states and this is
also the number of different positions that the
unoccupied site can have in the $s1$ effective
lattice.
 
Transitions between configuration states where the
unoccupied site has different positions in the $s1$
effective lattice result from a set of well-defined
elementary steps which are behind its motion around in 
the spin and $s1$ effective lattices. 
The generators of such processes have simple
and obvious expressions in terms of $s1$ bond-particle
creation and annihilation operators.
The above elementary steps correspond to 
changes in the occupancy configurations of rotated 
electrons and the unitary operator ${\hat{V}}$ preserves 
the occurrence of nearest hopping only for 
such objects. Therefore, in the case of the square lattice
rotated electrons can hop vertically or horizontally only
and given the relation between the original lattice
and the spin effective lattice that leads 
to restrictions in the changes of the position of
the link-site attachments. Indeed, the elementary
steps behind the net motion of the
unoccupied site involve horizontal or vertical
shifts by $\pm a_s$ of some of such attachments
where $a_s$ is the lattice constant of the
spin effective lattice given in Eq. (\ref{NNCC}).
The net length passed by the unoccupied
site in an elementary step is $2a_s$.

For 1D the unoccupied site moves by $\pm 2a_s$ in 
each elementary step. As for the square-lattice case, 
such elementary steps involve smaller virtual  
elementary steps where link-site attachments 
move by $\pm a_s$. For the square lattice the net
motion of the unoccupied site is given by
$\pm 2a_s\,\vec{e}_{x_i}$ where $i=1$ and  $i=2$ 
for horizontal and vertical elementary processes, respectively.
In such elementary steps the net length $2a_s$
passed by the $s1$ bond particle corresponds to
horizontal or vertical motion.
In addition, for the square lattice there are
horizontal-vertical or vertical-horizontal
elementary steps where the unoccupied
site moves by
$\pm a_s\,[\vec{e}_{x_i}+ \vec{e}_{x_{i'}}]$
or $\pm a_s\,[\vec{e}_{x_i}- \vec{e}_{x_{i'}}]$
where $i=1,2$ and $i'\neq i$. For the latter elementary 
steps the net length $2a_s$ corresponds to
both a horizontal and a vertical $\pm a_s$ shift,
so that the initial and final positions distance is
$\sqrt{2}\,a_s$

For the $g$-primary partitions any 
elementary process involves the virtual annihilation of 
one of the four $s1$ bond particles surrounding the
extra site. That virtual process involves the creation of two
unoccupied sites and is followed by suitable horizontal 
and/or vertical $\pm a_s$ shifts of well-defined 
link-site attachments. Those shifts are equivalent to 
the unoccupied site moving by
$\pm 2a_s\,\vec{e}_{x_i}$, $\pm a_s\,[\vec{e}_{x_i}+ \vec{e}_{x_{i'}}]$,
or $\pm a_s\,[\vec{e}_{x_i}- \vec{e}_{x_{i'}}]$
where $i=1,2$ and $i'\neq i$. Finally, a $s1$ bond particle is
created onto the site of the $s1$ effective lattice
unoccupied in the initial configuration so that two unoccupied 
sites of the virtual state are occupied.

For some $g$-primary partitions of the model on the square lattice 
such elementary steps also involve collective
processes where a whole kink row or anti-kink row
(and kink column or anti-kink column)
configuration moves from one row (and
column) to the nearest-neighboring 
or second-nearest-neighboring row
(and column). Such processes consist
of a collective horizontal (and vertical) 
shift of $-a_s$ or $+a_s$ of all the sites 
of a kink or anti-kink row 
(and kink or anti-kink column), respectively, plus
a collective horizontal (and vertical) opposite 
shift of $+a_s$ or $-a_s$ of all the sites of a 
nearest-neighboring or second-nearest-neighboring kink or 
anti-kink row (and kink or anti-kink column), 
respectively. 

Specifically, for $g$-primary partitions  
of horizontal (and vertical) sub-configuration states,
unoccupied-site horizontal (and vertical)
elementary steps do not involve such
kink or anti-kink row interchanges
(and kink or anti-kink column interchanges). 
In contrast, for $g$-primary partitions  
of vertical (and horizontal) sub-configuration states,
unoccupied-site horizontal (and vertical)
elementary steps involve kink or anti-kink column 
interchanges (and kink 
or anti-kink row interchanges) 
of second-nearest-neighboring 
columns (and rows). On the other hand,
horizontal-vertical or vertical-horizontal 
elementary steps involve 
kink or anti-kink row interchanges
of nearest-neighboring rows 
for $g$-primary partitions  
of horizontal sub-configuration states
and kink or anti-kink column interchanges 
of nearest-neighboring columns for primary
$g$-primary partitions  
of vertical sub-configuration states.

Any motion of the unoccupied site around in 
the spin effective lattice can be generated by a 
set of the above elementary steps. 
The processes behind such steps 
are easiest to describe for the primary partitions. However, they are also well-defined 
for the remaining partitions.

\subsection{Correspondence between the $N^h_1=1$ and
$N^h_1=0$ configuration states}

The partitions that span the $2D=2,4$ sub-configuration 
states of a $N_{s1}^h=1$ configuration state are independent
so that the number of partitions of the latter state is 
$2D=2,4$ times larger than that of the $N_{s1}^h=0$ 
configuration state. Indeed, there is an one-to-one 
correspondence between the partitions of that state  
and those of a $N_{s1}^h=1$ sub-configuration state
with the same number of $s1$ bond particles.
Concerning the primary partitions of
$N_{s1}^h=1$ sub-configuration states,
except for the sites of the spin effective lattice
belonging to the row and column 
which the unoccupied site belongs to, the real-space 
coordinates of the sites of that lattice
have the same real-space coordinates as those of the
corresponding primary partitions of the
$N^h_{s1}=0$ configuration state with
the same number of $s1$ bond particles. 

A $N^h_{s1}=1$ configuration state is defined by
the position of its extra site whose real-space coordinate 
in the $s1$ effective lattice is the same as 
that of the corresponding unoccupied site. 
Except for the sites belonging to the kink row and
anti-kink row (and kink column and anti-kink 
column), respectively, for the horizontal-kink and 
horizontal-anti-kink (and vertical-kink and vertical-anti-kink) 
sub-configuration state the two sites of each of the 
$N_{s1}$ links of the $2D=2,4$ primary partitions have 
the same real-space coordinates as the corresponding 
primary partitions of the $N^h_{s1}=0$ configuration 
state. 

The shifts of the real-space coordinates of the sites of the 
primary partitions of the $N_{s1}^h=0$ configuration state
which become sites of a $d,l$ line in the corresponding final 
sub-configuration state lead to the following real-space coordinates 
for the latter sites,
\begin{equation}
{\vec{r}}^{\,0}_j \rightarrow \vec{r}_j 
= {\vec{r}}^{\,0}_j + 2\vec{r}_{d,l}^{\,0} 
= {\vec{r}}^{\,0}_j + l\,a_s\,\vec{e}_d \, .
\label{trans-1h}
\end{equation}
Here ${\vec{r}}^{\,0}_j$ denotes the corresponding
real-space coordinate of the initial $N_{s1}^h=0$ configuration state.

The procedure to construct the non-primary partitions of 
a $N^h_{s1}=1$ sub-configuration state profits from all 
two-site links remaining exactly the same
and only the real-space coordinates of the sites of the 
spin effective lattice of the $N^h_{s1}=1$ 
sub-configuration state whose real-space coordinates 
are shifted for the primary partitions 
as given in Eq. (\ref{trans-1h}) being shifted exactly by
the same amount for the corresponding non-primary partitions. 
Such a procedure applies to any of the $N_{s1}+1$ positions
of the unoccupied site in the $s1$ effective lattice
and hence to all corresponding $N_{s1}^h=1$
configuration states. It corresponds to the elementary
processes generated by application of an one-electron
operator onto a $x\geq 0$ and $m=0$ ground state.

\subsection{$s1$ bond-particle operators of $N^h_1=1$ configuration states}

Each of the $2D=2,4$ sub-configuration states
of a $N^h_{s1}=1$ configuration state has
a $d,l$ line in the corresponding $d,l$ $g$-primary partition 
where for $1D$ the index $d$ is given only by $d=1$.
We use a single index $i\equiv d,l$ to denote
the indices $d,l$ which are used to label each 
sub-configuration state according to the type of $d,l$ line in its 
$g$-primary partition. For a $N^h_{s1}=1$ configuration 
state the operators given in Eq. (\ref{g-s1+general})
for the $N^h_{s1}=0$ configuration state read instead,
\begin{eqnarray}
g_{\vec{r}_{j},s1} & = & {1\over \sqrt{2D}}\sum_{i}g_{\vec{r}_{j},s1,i} \, ;
\hspace{0.15cm}
g_{\vec{r}_{j},s1}^{\dag} = \left(g_{{\vec{r}}_{j},s1}\right)^{\dag} 
\, ; \hspace{0.15cm}
g_{\vec{r}_{j},s1,i} = 
\sum_{g=0}^{N_{s1}/2D-1} h_{g,i}\, a_{\vec{r}_{j},s1,i,g}  \, , 
\nonumber \\
a_{\vec{r}_{j},s1,i,g} & = &
\sum_{d''=1}^{D}\sum_{l''=\pm1}
\, b_{\vec{r}_{j}+{\vec{r}_{d'',l''}}^{\,0},s1,i,d'',l'',g} 
\, ; \hspace{0.5cm}
\sum_{g=0}^{[N_{s1}/2D-1]} \vert h_{g,i}\vert^2 = {1\over 2D} \, ,
\label{g-s1+general-Nh-1}
\end{eqnarray}
where $D=1,2$ and the operator $g_{\vec{r}_{j},s1,i}$ acts onto
the spin and $s1$ effective lattice associated with
the $N^h_{s1}=1$ configuration state.
For each sub-configuration state the real-space 
coordinates $\vec{r}_{j}$ of the $s1$ bond particles
are the same as for the $N^h_{s1}=0$ configuration state
except for the corresponding $d,l$ line. For that
line they are shifted by $2\vec{r}_{d,l}^{\,0}$ as given
in Eq. (\ref{trans-1h}). In the above equation
we denote by $d''$ and $l''$ 
the summation indices in the expression for
the operator $a_{\vec{r}_{j},s1,i,g}$ given in
Eq. (\ref{g-s1+general-Nh-1}) to distinguish them
from the constant $d,l$ line indices $d$ and $l$. 

The two-site one-bond operators $b_{\vec{r},s1,i,d'',l'',g}$ 
and $b_{\vec{r},s1,i,d'',l'',g}^{\dag}$ involved
in Eq. (\ref{g-s1+general-Nh-1}) have for $s1$ bond particles whose
real-space coordinates do not belong to $d,l,l'$ lines 
the same expression for all $2D=2,4$ sub-configuration states 
and read,
\begin{equation}
b_{\vec{r},s1,i,d'',l'',g}^{\dag} = b_{\vec{r},s1,d'',l'',g}^{\dag} \, ,
\label{g-s-l-most}
\end{equation}
where $b_{\vec{r},s1,d'',l'',g}^{\dag}$ is given in Eq. (\ref{g-s-l}).
In turn, when $\vec{r}_{j}$ belongs to the $d,l,0\equiv d,l$ line
such operators are given by,
\begin{eqnarray}
b_{\vec{r},s1,i,d'',l'',g}^{\dag} & = & 
{(-1)^{d+d''}\over\sqrt{2}} 
(\left[{1\over 2}+s^z_{\vec{r}-\vec{r}^{\,g}_{d'',l''}}\right]
s^-_{\vec{r}+\vec{r}^{\,g}_{d'',l''}+\vec{r}_{i}} 
\nonumber \\
& - & \left[{1\over 2}+s^z_{\vec{r}+\vec{r}^{\,g}_{d'',l''}+\vec{r}_{i}}\right]
s^-_{\vec{r}-\vec{r}^{\,g}_{d'',l''}}) \, ,
\nonumber \\
\vec{r}_{i} & = & -\delta_{d'',\bar{d}}\,\,2\vec{r}_{d,l}^{\,0} \, ,
\label{g-s-l-h-line}
\end{eqnarray}
where we recall that $\bar{1}=2$ and $\bar{2}=1$. Finally,
for the $s1$ bond particles whose real-space coordinate
$\vec{r}_{j}$ belongs to the $d,l,l'$ lines such that
$l'=\pm 1$ the two-site one-bond operators read,
\begin{eqnarray}
b_{\vec{r},s1,i,d'',l'',g}^{\dag} & = & 
{(-1)^{d-1''}\over\sqrt{2}} 
(\left[{1\over 2}+s^z_{\vec{r}-\vec{r}^{\,g}_{d'',l''}}\right]
s^-_{\vec{r}+\vec{r}^{\,g}_{d'',l''}+\vec{r}_{i}} 
\nonumber \\
& - & \left[{1\over 2}+s^z_{\vec{r}+\vec{r}^{\,g}_{d'',l''}+\vec{r}_{i}}\right]
s^-_{\vec{r}-\vec{r}^{\,g}_{d'',l''}}) \, ,
\nonumber \\
\vec{r}_{i} & = & \delta_{d'',\bar{d}}\, \delta_{l'',-l'}
\,2\vec{r}_{d,l}^{\,0} \, .
\label{g-s-ll-h-line}
\end{eqnarray}

\subsection{More about $N^h_{s1}=0,1,2$ configuration states}

The $s1$ fermion momentum occupancies of Ref. \cite{companion} of excited states 
generated by application onto $x\geq 0$ and $m=0$
ground states of two-electron operators involving (i) creation
or annihilation of two electrons with the same spin
projection and (ii) spin-triplet and spin-singlet excitations
are described by a superposition of the set of
$[N_{s1}+2]\,[N_{s1}+1]/2$ configuration states with
$N^h_{s1}=2$ unoccupied sites studied in the following. 
Indeed, there are $[N_{s1}+2]\,[N_{s1}+1]/2$ different $N^h_{s1}=2$ 
configuration states corresponding to the possible 
$[N_{s1}+2]\,[N_{s1}+1]/2$ positions of the two unoccupied 
sites in the $s1$ effective lattice. According to Eq. (\ref{Nas1-Nhs1}),
here $N_{s1}+2 = [N_{a_s}^D/2 + S_s]$ where $S_s=1$
for $N_{s2}=0$ and $S_s=0$ for $N_{s2}=1$ is the number 
of sites $N_{a_{s1}}^D$ of that lattice whose 
number of unoccupied sites reads $N^h_{s1} =  [2S_s +2N_{s2}] =2$. 
For such states the number 
$N_{a_s}^D=[2N_{s1}+4-2S_s]$ of sites of the spin effective 
lattice is even. There are two sets of 
$[N_{s1}+2]\,[N_{s1}+1]/2$ configuration states
which contribute to different $N^h_{s1}=2$ energy
eigenstates. Those are the $N^h_{s1}=2$, $S_s=1$,
$N_{s2}=0$ spin-triplet configuration states
and $N^h_{s1}=2$, $S_s=0$, $N_{s2}=1$ 
spin-singlet configuration states, respectively.

Since it is assumed that for
the square lattice $[2N_{s1}]^{1/2}$ is an integer 
number, the notations $N_{a_{s1}}^2$ and $N_{a_s}^2$ 
do not mean that the numbers $N_{a_{s1}}$ and 
$N_{a_s}$ are integers. $N_{a_{s1}}^2=N_{s1}+2$ and 
$N_{a_s}^2=[2N_{s1}+4-2S_s]$ are integer numbers. In turn, the integer 
numbers closest to $N_{a_{s1}}$ and $N_{a_s}$ are the mean
value of the number of sites of each row and column
of the $s1$ and spin effective lattice, respectively. 
Our study takes into account the exact
number of sites of such rows and columns.

Concerning the above two types of $N_{s1}^h=2$ configuration 
states, the two unoccupied sites 
can either correspond to two independent $+1/2$ spinons ($S_s=1$
and $N_{s2}=0$) or two spinons out of the four which are 
part of the $s2$ fermion ($S_s=0$ and $N_{s2}=1$). 
We recall that alike an independent spinon, the spin-neutral
four-spinon occupancy configuration of such a $s2$ fermion remains invariant under
the electron - rotated-electron unitary transformation \cite{companion}.

For the $N_{a_{s}}^2\rightarrow\infty$ limit considered here 
for the model on the square lattice the 
correct physics is achieved if one considers that out of the 
$[2N_{s1}+4-2S_s]$ sites of the spin effective lattice, $2N_{s1}$ sites 
correspond to a square lattice and the position of the two extra 
sites which contribute to the $s1$ effective lattice
is defined as for the $N^h_{s1}=1$ configuration states. 
For the model both on the 1D and square lattice and $S_s=0$ and thus $N_{s2}=1$ the number of
sites of the spin effective lattice is given by
$N_{a_s}^D=2N_{s1}+4$. However, two of the four extra sites do not 
contribute to the $s1$ effective lattice. The point is that 
for $N^h_{s1}=2$ configuration states with $S_s=0$ and $N_{s2}=1$
when the $s1$ bond particles moves around in that lattice, 
they use as unoccupied sites only two out of the four 
sites of the $s2$ fermion. Two of these four sites of
the spin effective lattice do not belong to the $s1$ effective 
lattice, alike the sites of the $\eta$-spin effective lattice 
associated with the original rotated-electron unoccupied sites
do not belong to the spin effective lattice. 

The spin $SU(2)$ symmetry imposes that the $s1$ effective lattices 
of the $N^h_2=2$ configuration states such that
(i) $S_s=1$ for $N_{s2}=0$ and (ii) $S_s=0$ for $N_{s2}=1$,
respectively, must be identical. In either case each of the 
two extra sites obeys
the boundary conditions given in Eq. (\ref{x-0-d}) which
for the present $N_{a_s}\rightarrow\infty$ limit yield the
correct physics. For the square lattice the row and column 
periodic boundary conditions are similar to those of the
$N^h_{s1}=1$ configuration states. For the 1D lattice periodic 
conditions are also used. 

Symmetry imposes that for $g$-primary partitions of 
$N^h_{s1}=2$ configuration states 
the two unoccupied sites are associated with 
kink-like and anti-kink-like link configurations, 
respectively. Moreover, symmetry implies that only 
sub-configuration states for whose 
$g$-primary partitions both the kink and anti-kink 
link configurations associated with the two unoccupied sites,
respectively, correspond to rows or columns are allowed. 
It follows that alike for $N^h_{s1}=1$ configuration states, a
$N^h_{s1}=2$ configuration state involves $2D=2,4$ 
sub-configuration states. In the present case
there are two horizontal sub-configuration
states for whose $g$-primary partitions
the unoccupied sites of real-space coordinates 
$\vec{r}_{j_0}$ and $\vec{r}_{j_0'}$ are associated with
row kink-like and anti-kink-like link configurations and row 
anti-kink-like and kink-like link configurations, respectively. 
Furthermore, there are two vertical 
sub-configuration states for whose 
$g$-primary partitions the unoccupied sites 
with such real-space coordinates are associated with 
column kink-like and anti-kink-like link configurations and 
column anti-kink-like and kink-like link configurations, respectively. 

It is useful to classify the $N^h_{s1}=2$ configuration 
states into three main classes:
\vspace{0.25cm}

1) $N^h_{s1}=2$ configuration states whose two 
extra sites do not belong to the same row and column. 
Such states occur for the model on the square lattice but do not exist 
for the 1D lattice. For the corresponding $g$-primary partitions the occupied sites are
the extra sites, which correspond to kink-like or anti-kink-like
link configurations of the same form as those of
$N^h_{s1}=1$ states. Indeed, each unoccupied 
site behaves independently
and can move around in the spin effective lattice 
by means of the elementary steps discussed above
for the $N^h_{s1}=1$ configuration states. 
For all $g$-primary partitions one unoccupied site is associated with
a kink-like link configuration and the other with an anti-kink-like
link configuration, respectively. The $s1$
bond-particle operators are defined as for the 
$N^h_{s1}=1$ configuration states by Eqs.
(\ref{g-s1+general-Nh-1})-(\ref{g-s-ll-h-line})
but with parallel $d,+1$ and $d,-1$ lines and the
index $i$ corresponding to the four sub-configuration
states defined above for the $N^h_{s1}=2$ configuration 
states.
\vspace{0.25cm}

2) $N^h_{s1}=2$ configuration states whose two extra 
sites belong to the same row or column but are not nearest 
neighboring sites. For the 1D lattice only states of this
type and of type 3 exist. Let the two extra sites
have real-space coordinates 
$\vec{r}_{j_0}$ and $\vec{r}_{j_0'}$. Then for the
$g$-primary partition 
of one of the two horizontal (and vertical) sub-configuration 
states the link occupancy involving sites belonging
to the row part (and column part) located 
between the sites of real-space
coordinates $\vec{r}_{j_0}$ and $\vec{r}_{j_0'}$
has opposite sign relative to those of the surrounding rows 
(and columns) whereas the link occupancy 
involving sites belonging
to the row part (and column part) located between 
the sites of real-space coordinates $\vec{r}_{j_0'}$ and $\vec{r}_{j_0}$
has the same sign as those of the surrounding rows.
In contrast, for the $g$-primary partition 
of the other horizontal (and vertical) sub-configuration 
states the sign of the link occupancies on the
two above row parts (and column parts) is the opposite. 
For $g$-primary partitions 
the two unoccupied sites of real-space
coordinates $\vec{r}_{j_0}$ and $\vec{r}_{j_0'}$ 
behave as kink or anti-kink mobile domain walls 
whose link structure is shown in Figs. 7 and 9. The $s1$
bond-particle operators are defined as for the 
$N^h_{s1}=1$ configuration states by Eqs.
(\ref{g-s1+general-Nh-1})-(\ref{g-s-ll-h-line})
but with the operators whose real-space coordinates
correspond to link occupancy configurations with
opposite sign relative to those of the surrounding rows or 
columns referring to the $d,j$ line parts defined
above instead of to the whole $d,j$ line.
\vspace{0.25cm}

3) $N^h_{s1}=2$ configuration states whose two extra
sites belong to the same row and column and are nearest-neighboring 
sites in the spin effective lattice. In that case there
is both a horizontal and a vertical 
sub-configuration state whose extra-site row and
column link occupancies have opposite sign relative
to those of the surrounding rows and columns, respectively.
In turn, for the other horizontal and vertical 
sub-configuration states, respectively, all link occupancies have the same
sign and the two-unoccupied-site link occupancy configuration 
of the corresponding $g$-primary partitions 
are identical. A small sub-domain of such a link configuration 
including the two-unoccupied sites of the spin effective
lattice is shown on the right-hand side of Fig. 4. These
configuration states can be generated from those of
type 2. This is achieved by moving both unoccupied sites through
their row or column, one against the other, until they 
become nearest-neighbor sites. For the square (and 1D) lattice,
the $g$-primary partitions of two (and one) out of the four 
(and two) sub-configuration states
such a motion and resulting "collision" leads to the full cancellation of the 
corresponding kink-like and anti-kink-like link occupancy configurations. 
Such a unoccupied-site kink and anti-kink annihilation 
has similarities with a particle - anti-particle annihilation
process. The corresponding "collisions" occur more often for
a unoccupied site moving around in the spin effective
lattice of the 1D lattice. Indeed, then there is a single
chain. In contrast, for the model on the square lattice there are
both horizontal and vertical elementary steps so
that the two unoccupied sites can move independently
around in the spin effective lattice without colliding provided
that their motion involves configuration states
of type 1, which do not exist in 1D. For the two 
above sub-configuration states of the
model on the square lattice whose extra-site row or
column link occupancies have opposite sign relative
to those of the surrounding rows or columns, respectively, 
the $s1$ bond-particle operators are defined as for the 
$N^h_{s1}=1$ configuration states by Eqs. (\ref{g-s1+general-Nh-1})-(\ref{g-s-ll-h-line}).
For the other two sub-configuration 
states whose extra-site row or
column link occupancies have the same sign as those of 
the surrounding rows or columns, respectively, 
the $s1$ bond-particle operators are defined alike for the 
$N^h_{s1}=0$ configuration states by Eq. (\ref{g-s1+general}).
\vspace{0.25cm}

The number $N^h_{s1}$ of unoccupied sites is a good
quantum number for the model on the square lattice in the one- and
two-electron subspace so that the $s1$ fermion momentum occupancies of Ref. \cite{companion}
of the states that span such a subspace 
are described by a suitable superposition of configuration states
with constant number of unoccupied sites in the $s1$
effective lattice. For instance, the $s1$ fermion momentum
occupancy of a $N^h_{s1}=2$ spin-triplet (and spin-singlet)
excited state is described by a suitable superposition of
the $[N_{s1}+2]\,[N_{s1}+1]/2$ $N^h_{s1}=2$ configuration 
states with $S_s=1$ and $N_{s2}=0$
(and with $S_s=0$ and $N_{s2}=1$.) Application of 
a $s1$ bond-particle creation operator of
real-space coordinate $\vec{r}_j$ onto such a ground state 
gives zero for configuration states of the above types 1
and 2: only the two sub-configuration states of a
configuration state of type 3 whose link configurations 
have all the same sign and the real-space coordinate of the 
central unoccupied site on the right-hand side of Fig. 4
coincides with that of the applied
operator are transformed onto the $N^h_{s1}=0$ configuration 
state with one more $s1$ bond particle than the
initial state. 

Moreover, application of a $s1$ bond-particle creation 
operator onto $N^h_{s1}=0,1$ configuration states gives 
always zero. In turn, application of a $s1$ bond-particle 
annihilation operator whose real-space coordinate is that 
of an occupied site of the $s1$ effective lattice of
a $N^h_{s1}=0,1,2$ configuration state transforms it 
into a state with two more unoccupied sites than the initial state.

Finally, in Appendix A it is confirmed that within the description 
used in the studies of Ref. \cite{companion} one has that for $U/4t>0$ and the 
subspaces where the $s1$ bond-particle operators of the 
one- and two-electron and $N_{s1}^h=0,1,2$
subspace are defined, upon acting onto the $s1$ effective lattice 
such operators anticommute on the same site of that lattice,
\begin{equation}
\{g^{\dag}_{{\vec{r}}_{j},s1},g_{{\vec{r}}_{j},s1}\} = 1 \, ;
\hspace{0.15cm}
\{g^{\dag}_{{\vec{r}}_{j},s1},g^{\dag}_{{\vec{r}}_{j},s1}\} =
\{g_{{\vec{r}}_{j},\alpha\nu},g_{{\vec{r}}_{j},s1}\}=0 \, ,
\label{g-local}
\end{equation}
and commute on different sites,
\begin{equation}
[g^{\dag}_{{\vec{r}}_{j},s1},g_{{\vec{r}}_{j'},s1}] =
[g^{\dag}_{{\vec{r}}_{j},s1},g^{\dag}_{{\vec{r}}_{j'},s1}]
= [g_{{\vec{r}}_{j},s1},g_{{\vec{r}}_{j'},1}] = 0 \, ,
\label{g-non-local}
\end{equation}
where $j\neq j'$. That algebra confirms that the $s1$ bond-particle 
operators are hard-core like. 

\section{Concluding remarks}

For the Hubbard model on the square lattice in the one- and two-electron subspace the number 
$N^h_{s1}$ of unoccupied sites of the $s1$ effective lattice is a good quantum number. Such a 
subspace is spanned by the states with $N^h_{s1}=0,1,2$, whose spin configurations are 
investigated in this paper. In the studies of Sections 3 and 4 a change of gauge 
structure \cite{Xiao-Gang} is considered so that the real-space coordinates of the sites of the 
$s1$ effective lattice correspond to one of the two 
sub-lattices of the spin effective lattice. That change is
fulfilled for the $N^h_{s1}=0$ configuration state for which
the gauge structure occurs. That leads to two alternative
definitions of the $s1$ effective lattice for both
that state and the related $N^h_{s1}=1,2$ configuration
states: As discussed in this paper, the occupancy configurations of the 
latter states can be generated from those of the 
$N^h_{s1}=0$ configuration state with the same 
number of $s1$ bond particles.

The $N^h_{s1}=0$ configuration states studied here generate
the spin degrees of freedom of the $x\geq 0$ and $m=0$ ground states and their 
two-electron charge excited states of the Hubbard model on the square lattice 
in the one- and two-electron subspace. That quantum problem refers to the square-lattice
quantum liquid of $c$ and $s1$ fermions of Ref. \cite{companion}. In turn, the spin
configurations of the excited states generated from
application onto these ground states of one-electron operators and two-electron
operators other than the charge operator are suitable superpositions of the $N^h_{s1}=1$   
and $N^h_{s1}=0$ or $N^h_{s1}=2$ configuration states,
respectively, studied in this paper.
For the $N^h_{s1}=1,2$ states the unoccupied sites move
around in the spin effective lattice through well-defined elementary
steps. For $N^h_{s1}=2$ 
configuration states the two unoccupied sites may move
through the same row or column, one against the other, 
until they become nearest-neighbor sites. For the model on the
square (and 1D) lattice, for two out of four (and one out of two) 
$g$-primary partitions defined in this paper such a motion 
and resulting "collision" leads to the full cancellation of the 
corresponding kink-like and anti-kink-like link occupancy configurations,
similarly to a particle - anti-particle annihilation
process. Such collisions occur more often for
an unoccupied site moving around in the spin effective
lattice of the model on the 1D lattice, which corresponds to a single chain.
As confirmed in Appendix A,
the spin configurations obtained from such a cancellation
play an important role in the $s1$ bond-particle operator
algebra.

Moreover, in this paper suitable $s1$ bond-particle 
operators have been constructed, which 
upon acting onto the $s1$ effective lattice obey a hard-core 
like algebra. Such a property is valid for the subspaces
in which the $s1$ bond-particle operators act onto. It plays an
important role in the related studies of Ref. \cite{companion}.
There corresponding $s1$ fermion operators are generated
by means of an extended Jordan-Wigner transformation from 
the $s1$ bond-particle operators introduced in this paper.
Concerning previous studies on the large-$U$ Hubbard
model and $t-J$ model on a square lattice involving for instance
the slave particle formalism \cite{2D-MIT,Fazekas,Xiao-Gang} or Jordan-Wigner 
transformations \cite{Feng}, the crucial requirement is to impose
the single occupancy constraint. Here that constraint is 
naturally implemented for all values of $U/4t>0$, since the
spins associated with the spin $SU(2)$ state representations
refer to the rotated electrons of the singly occupied sites.
Moreover, for the above schemes the spinless fermions arise from 
individual spin-$1/2$ spins or spinons. In contrast, within the extended
Jordan-Wigner transformation
performed in Ref. \cite{companion}
the $s1$ fermions emerge from the spin-neutral two-spinon composite 
$s1$ bond particles studied in this paper.

The studies of Ref. \cite{companion} confirm that the results
obtained here concerning the Hubbard model on the square lattice in the one- and 
two-electron subspace are useful for the further understanding of the role of
electronic correlations in the spin-wave spectrum observed in the parent compound 
La$_2$CuO$_4$ \cite{LCO-neutr-scatt}. A system of weakly coupled planes, each described 
by the square-lattice quantum liquid of Ref. \cite{companion}, perturbed by the
effects of intrinsic disorder is expected to be the simplest realistic toy model for the 
description of the role of correlations effects in the 
unusual properties of the cuprate hight-temperature superconductors  
\cite{two-gaps,k-r-spaces,2D-MIT,ARPES-review}.

I thank Nuno M. R. Peres for discussions and support in the figures production. 
I also thank Miguel A. N. Ara\'ujo, Daniel Arovas, Pedro D. Sacramento, and 
Maria J. Sampaio for discussions and the support of the ESF Science 
Program INSTANS and grant PTDC/FIS/64926/2006.

\appendix

\section{Hard-core character of the $s1$ bond-particle operators}

The goal of this Appendix is to confirm the validity of the relations 
provided in Eqs. (\ref{g-local}) and (\ref{g-non-local}).
In order to probe such relations, we consider without any loss of generality 
that the initial state is a $N^h_{s1}=4$ configuration state.
Indeed and as discussed in Subsection 4-5, application of a $s1$ bond-particle creation 
operator onto $N^h_{s1}=0,1$ configuration states gives 
zero. Application of two $s1$ bond-particle creation 
operators onto $N^h_{s1}=2$ configuration states gives 
zero as well. The studies on $N^h_{s1}=2$ configuration
states of that subsection can be straightforwardly
generalized to $N^h_{s1}=4$ configuration states. 
That assures that application onto those of four-site
two-bond operator may not give zero. Fortunately, the final results 
reached here apply to any configuration state with a finite 
number $N^h_{s1}$ of unoccupied sites in the $s1$
effective lattice, yet are simpler to derive for the
$N^h_{s1}=4$ configuration states.

Sub-configuration states of type 3 considered in 
Subsection 4-5 for $N^h_{s1}=2$ 
also exist for $N^h_{s1}=4$. Within those with
$N^h_{s1}=4$ unoccupied sites,
the importance of sub-configuration states 
whose link configurations have all the same sign 
stems from those being the only ones for which application of
$s1$ bond-particle creation operators does not give
zero. For other $N^h_{s1}$-finite sub-configuration 
states the $s1$ bond-particle operators have slightly
more involved expressions given in Section 4. Fortunately, the use 
of the operator expressions suitable
for sub-configuration states of type 3
leads to the same final results. Therefore, for simplicity
we consider here the latter $s1$ bond-particle operators 
whose expressions are the same as for 
the $N^h_{s1}=0$ configuration state studied in
Section 3.

All the results of this Appendix apply to the model on both the square
and 1D lattices. The 1D expressions are readily obtained if one considers
in the general expressions given below only $d=1$ contributions and terms,
together with the choice $D=1$ in the $D$-dependent quantities. 
In order to reach our goal, let us recall that the rotated-electron operators 
are related to the original electron operators as follows,
\begin{equation}
{\tilde{c}}_{\vec{r}_j,\sigma}^{\dag} =
{\hat{V}}^{\dag}\,c_{\vec{r}_j,\sigma}^{\dag}\,{\hat{V}} \, ,
\label{tilde-c-c}
\end{equation}
where ${\hat{V}}$ is the electron -
rotated-electron unitary operator. 
Straightforward manipulations based on Eqs.
(\ref{fc+})-(\ref{rotated-quasi-spin}) then lead
to the following algebra for the related $c$ fermion operators \cite{companion},
\begin{equation}
\{f_{\vec{r}_j,c}\, ,f_{\vec{r}_{j'},c}^{\dag}\} = \delta_{j,j'} 
\, ; \hspace{0.5cm}
\{f_{\vec{r}_j,c}^{\dag}\, ,f_{\vec{r}_{j'},c}^{\dag}\} =
\{f_{\vec{r}_j,c}\, ,f_{\vec{r}_{j'},c}\} = 0 \, ,
\label{albegra-cf}
\end{equation}
$c$ fermion operators and rotated-quasi-spin operators,
\begin{equation}
[f_{\vec{r}_j,c}^{\dag}\, ,q^l_{\vec{r}_{j'}}] =
[ f_{\vec{r}_j,c}\, ,q^l_{\vec{r}_{j'}}] = 0 \, ,
\label{albegra-cf-s-h}
\end{equation}
and rotated-quasi-spin operators,
\begin{equation}
[q^{x_p}_{\vec{r}_j}\, ,q^{x_{p'}}_{\vec{r}_{j'}}] =
i\,\delta_{j,j'}\sum_{p''} \epsilon_{pp'p''}\,q^{p''}_{\vec{r}_j} 
\, ; \hspace{0.15cm} p=1,2,3 \, ,
\label{albegra-s-h}
\end{equation}
\begin{equation}
\{q^{+}_{\vec{r}_j},q^{-}_{\vec{r}_j}\} = 1 \, ,
\hspace{0.5cm}
\{q^{\pm}_{\vec{r}_j},q^{\pm}_{\vec{r}_j}\} = 0 \, ,
\label{albegra-qs-p-m}
\end{equation}
\begin{equation}
[q^{+}_{\vec{r}_j},q^{-}_{\vec{r}_{j'}}] = \delta_{j,j'}\,2q^{x_3}_{\vec{r}_j}
\, , \hspace{0.5cm}
[q^{\pm}_{\vec{r}_j},q^{\pm}_{\vec{r}_{j'}}] = 0 \ \, .
\label{albegra-q-com}
\end{equation}
Hence the rotated-quasi-spin operators $q^{\pm}_{\vec{r}_j}$ anticommute 
on the same site and commute on different sites.

Moreover, combining the 
expressions given in Eq. (\ref{sir-pir}) with the algebraic relations
provided in Eqs. (\ref{albegra-s-h})-(\ref{albegra-q-com}) one readily finds that,
\begin{equation}
\{s^{+}_{\vec{r}_j},s^{-}_{\vec{r}_j}\} = 1 \, ,
\hspace{0.5cm}
\{s^{\pm}_{\vec{r}_j},s^{\pm}_{\vec{r}_j}\} = 0 \, ,
\label{albegra-s-p-m}
\end{equation}
\begin{equation}
[s^{+}_{\vec{r}_j},s^{-}_{\vec{r}_{j'}}] =
[s^{\pm}_{\vec{r}_j},s^{\pm}_{\vec{r}_{j'}}]=0 \, ,
\label{albegra-s-com}
\end{equation}
for $j\neq j'$ and,
\begin{equation}
[s^{z}_{\vec{r}_j},s^{\pm}_{\vec{r}_{j'}}] =
\pm\delta_{j,j'}s^{\pm}_{\vec{r}_{j}} \, .
\label{albegra-s-sz-com}
\end{equation}
It follows that the spinon operators $s^{\pm}_{\vec{r}_j}$ anticommute 
on the same site and commute on different sites. Consistently 
with the rotated-electron singly-occupied site projector 
$n_{\vec{r}_j,c}$ appearing in the expression of the spinon 
operators $s^{\pm}_{\vec{r}_j}$ and $s^z_{\vec{r}_j}$
provided in Eq. (\ref{sir-pir}), their real-space coordinates 
$\vec{r}_j$ can in the present $N_a^D\rightarrow\infty$ limit be identified
with those of the spin effective lattice. Therefore,
the corresponding operator index values $j=1,...,N_{a_s}$
are chosen to be those of that lattice.

For the sub-configuration states of type 3
the operators $g^{\dag}_{{\vec{r}}_{j},s1}$ (and $g_{{\vec{r}}_{j},s1}$) 
which create (and annihilate) a $s1$ bond particle at a site of 
the spin effective lattice of real-space coordinate ${\vec{r}}_{j}$ have 
the general form given in Eq. (\ref{g-s1+general}) both for the 
model on the 1D and square lattices,
where the absolute value $\vert h_{g}\vert$
of the coefficients $h_{g}$ decreases for increasing link length 
$\xi_{g}$ and obeys the normalization sum-rule
(\ref{g-s1+sum-rule}). Hence the expression of 
the $s1$ bond-particle operator $g_{\vec{r}_{j},s1}^{\dag} $ involves
the operators $a_{\vec{r}_{j},s1,g}^{\dag}$ and $a_{\vec{r}_{j},s1,g}$ 
of Eq. (\ref{g-s1+general}), which 
create and annihilate, respectively, a superposition of $2D=2,4$ two-site bonds of 
the same type and $b_{\vec{r},s1,d,l,g}^{\dag}$ and $b_{\vec{r},s1,d,l,g}$ are
two-site one-bond operators whose expression is given in Eq. (\ref{g-s-l}).

In order to confirm the validity of Eqs. (\ref{g-local}) and (\ref{g-non-local})
let us use Eqs. (\ref{g-s1+general})-(\ref{g-s-l}) to rewrite the anti-commutation 
relations of Eq. (\ref{g-local}) in terms of anti-commutators of two-site one-bond 
operators as follows,
\begin{equation}
\{g^{\dag}_{{\vec{r}}_{j},s1},g_{{\vec{r}}_{j},s1}\} = 
\sum_{d,l,g}\sum_{d',l',g'} h^*_g\,h_{g'}
\{b_{{\vec{r}}_j +\vec{r}_{d,l}^{\,0},s1,d,l,g}^{\dag},
b_{{\vec{r}}_j +\vec{r}_{d',l'}^{\,0},s1,d',l',g'}\} \, ,
\label{g-local-AP+-}
\end{equation}
\begin{equation}
\{g^{\dag}_{{\vec{r}}_{j},s1},g^{\dag}_{{\vec{r}}_{j},s1}\} = 
\sum_{d,l,g}\sum_{d',l',g'} h^*_g\,h^*_{g'}
\{b_{{\vec{r}}_j +\vec{r}_{d,l}^{\,0},s1,d,l,g}^{\dag},
b_{{\vec{r}}_j +\vec{r}_{d',l'}^{\,0},s1,d',l',g'}^{\dag}\} \, ,
\label{g-local-AP++}
\end{equation}
\begin{equation}
\{g_{{\vec{r}}_{j},s1},g_{{\vec{r}}_{j},s1}\} = 
\sum_{d,l,g}\sum_{d',l',g'} h_g\,h_{g'}
\{b_{{\vec{r}}_j +\vec{r}_{d,l}^{\,0},s1,d,l,g},
b_{{\vec{r}}_j +\vec{r}_{d',l'}^{\,0},s1,d',l',g'}\} \, ,
\label{g-local-AP--}
\end{equation}
where for simplicity we used the abbreviated summation
notation,
\begin{equation}
\sum_{d,l,g} \equiv \sum_{d=1}^{D}\sum_{l=\pm 1}\sum_{g=0}^{N_{s1}/2D-1} \, .
\label{sum}
\end{equation}
Moreover, by the use of the same equations the commutation relations of 
Eq. (\ref{g-non-local}) can be expressed in terms commutators of two-site 
one-bond operators. That leads to,
\begin{equation}
[g^{\dag}_{{\vec{r}}_{j},s1},g_{{\vec{r}}_{j'},s1}] = 
\sum_{d,l,g}\sum_{d',l',g'} h^*_g\,h_{g'}
[b_{{\vec{r}}_j +\vec{r}_{d,l}^{\,0},s1,d,l,g}^{\dag},
b_{{\vec{r}}_{j'} +\vec{r}_{d',l'}^{\,0},s1,d',l',g'}] \, ,
\label{g-non-local-AP+-}
\end{equation}
\begin{equation}
[g^{\dag}_{{\vec{r}}_{j},s1},g^{\dag}_{{\vec{r}}_{j'},s1}] = 
\sum_{d,l,g}\sum_{d',l',g'} h^*_g\,h^*_{g'}
[b_{{\vec{r}}_j +\vec{r}_{d,l}^{\,0},s1,d,l,g}^{\dag},
b_{{\vec{r}}_{j'} +\vec{r}_{d',l'}^{\,0},s1,d',l',g'}^{\dag}] \, ,
\label{g-non-local-AP++}
\end{equation}
\begin{equation}
[g_{{\vec{r}}_{j},s1},g_{{\vec{r}}_{j'},s1}] = 
\sum_{d,l,g}\sum_{d',l',g'} h_g\,h_{g'}
[b_{{\vec{r}}_j +\vec{r}_{d,l}^{\,0},s1,d,l,g},
b_{{\vec{r}}_{j'} +\vec{r}_{d',l'}^{\,0},s1,d',l',g'}] \, ,
\label{g-non-local-AP--}
\end{equation}
where $j\neq j'$.

According to the studies of Subsection 3-3, four rules follow 
from the definition of the subspace where the operators of Eqs. 
(\ref{g-s1+general})-(\ref{g-s-l}) act onto.
The evaluation of the anti-commutators and commutators
of the two-site one-bond operators on the right-hand side
of Eqs. (\ref{g-local-AP+-})-(\ref{g-local-AP--}) and
(\ref{g-non-local-AP+-})-(\ref{g-non-local-AP--}),
respectively, relies on both such rules and the algebra 
given in Eqs. (\ref{albegra-s-p-m})-(\ref{albegra-s-sz-com})
of the spinon operators
$s^{\pm}_{\vec{r}_j}$ and $s^z_{\vec{r}_j}$ of Eq. (\ref{sir-pir}),
which are the building blocks of the two-site
one-bond operators of Eq. (\ref{g-s-l}).
Fortunately, according to Eqs. (\ref{albegra-s-p-m})-(\ref{albegra-s-sz-com})
the spinon operators $s^{\pm}_{\vec{r}_j}$ obey the
usual algebra: They anticommute 
on the same site of the spin effective lattice
and commute on different sites. 

The two-site one-bond operators of Eq. (\ref{g-s-l}) can be rewritten as,
\begin{equation}
b_{\vec{r}_1,\vec{r}_2}^{\dag} = 
{(-1)^{d-1}\over\sqrt{2}}\left(\left[{1\over 2}+s^z_{\vec{r}_1}\right]
s^-_{\vec{r}_2} - \left[{1\over 2}+s^z_{\vec{r}_2}\right]
s^-_{\vec{r}_1}\right) \, ,
\label{g-s-l-simple}
\end{equation}
and $b_{\vec{r}_1,\vec{r}_2} = \left(b_{\vec{r}_1,\vec{r}_2}^{\dag}\right)^{\dag}$
where recalling that the real-space coordinate of their link
centre reads $\vec{r} = \vec{r}_j+\vec{r}_{d,l}^{\,0}$ the
real-space coordinates $\vec{r}_1$ and $\vec{r}_2$ are
given by,
\begin{equation}
\vec{r}_1 = \vec{r}_j+\vec{r}_{d,l}^{\,0}-\vec{r}_{d,l}^{\,g} \, ;
\hspace{0.15cm}
\vec{r}_2 = \vec{r}_{j'}+\vec{r}_{d',l'}^{\,0}+\vec{r}_{d',l'}^{\,g'} \, .
\label{variables}
\end{equation}

The evaluation of the anti-commutators and commutators
of the two-site one-bond operators on the right-hand side
of Eqs. (\ref{g-local-AP+-})-(\ref{g-local-AP--}) and
(\ref{g-non-local-AP+-})-(\ref{g-non-local-AP--}),
respectively, then relies on straightforward manipulations based on Eqs. 
(\ref{albegra-s-p-m})-(\ref{albegra-s-sz-com}) and on the 
four rules given in Subsection 3-3, which define the subspace that the operator
algebra under consideration refers to. For the two general
anti-commutators needed to evaluate the two-site one-bond operators 
on the right-hand side of Eq. (\ref{g-local-AP+-}) we find the
following expressions,
\begin{eqnarray}
\{b_{\vec{r}_1,\vec{r}_2}^{\dag},b_{\vec{r}_1,\vec{r}_2}\} & = &
\sum_{i=1,2}\{{1\over 2}\left({1\over 2}+s^z_{\vec{r}_i}\right)^2 -
s^+_{\vec{r}_i}s^-_{\vec{r}_{\bar{i}}}\left({1\over 2}+s^z_{\vec{r}_i}\right)
\left({1\over 2}+s^z_{\vec{r}_{\bar{i}}}\right) 
\nonumber \\
& + & {1\over 2}s^+_{\vec{r}_i}s^-_{\vec{r}_{\bar{i}}}
\left[\left({1\over 2}+s^z_{\vec{r}_i}\right) -
\left({1\over 2}+s^z_{\vec{r}_{\bar{i}}}\right)\right]\} \, ,
\label{anti-com-b+b-on}
\end{eqnarray}
\begin{eqnarray}
\{b_{\vec{r}_1,\vec{r}_2}^{\dag},b_{\vec{r}_{1'},\vec{r}_{2'}}\} & = & (-1)^{d+d'}
\sum_{i=1,2}\{s^+_{\vec{r}_{i'}}s^-_{\vec{r}_{i}}\left({1\over 2}+s^z_{\vec{r}_{\bar{i}'}}\right)
\left({1\over 2}+s^z_{\vec{r}_{\bar{i}}}\right)
\nonumber \\
& - & s^+_{\vec{r}_{i'}}s^-_{\vec{r}_{\bar{i}}}\left({1\over 2}+s^z_{\vec{r}_{\bar{i}'}}\right)
\left({1\over 2}+s^z_{\vec{r}_{i}}\right)\} \, ,
\label{anti-com-b+b-off}
\end{eqnarray}
where $\bar{1}=2$, $\bar{2}=1$, $\vec{r}_1\neq \vec{r}_{1'}, \vec{r}_{2'}$, and
$\vec{r}_2\neq \vec{r}_{1'}, \vec{r}_{2'}$. Indeed, according to
the second rule for the subspace where the two-site one-bond operators act onto only
general operators $b_{\vec{r}_1,\vec{r}_2}^{\dag}b_{\vec{r}_{1'},\vec{r}_{2'}}$
which do not join sites or join both sites of the spin effective
lattice lead to wanted and physical spin configurations. 

Moreover, in the initial configuration that the four-site two-bond
operators of the relation (\ref{anti-com-b+b-on}) act onto 
one has according to the first and third rules 
that the two sites of real-space coordinates $\vec{r}_1$
and $\vec{r}_2$ either (i) are occupied by two independent $+1/2$ spinons or
(ii) are linked by a bond. In turn, the
anti-commutator of Eq. (\ref{anti-com-b+b-off}) is of the
form of those on the right-hand side of
Eq. (\ref{g-local-AP+-}) so that in Eq. (\ref{variables}) one
has that $j=j'$ yet $\vec{r}_{d,l}^{\,g}\neq \vec{r}_{d',l'}^{\,g'}$
and the restrictions imposed by the forth rule 
must be taken into account. Since we find that
$b_{\vec{r}_1,\vec{r}_2}^{\dag}b_{\vec{r}_{1'},\vec{r}_{2'}}=
b_{\vec{r}_{1'},\vec{r}_{2'}}b_{\vec{r}_1,\vec{r}_2}^{\dag}$ 
and each of such operators is given by one half the
operator on the right-hand side of Eq. (\ref{anti-com-b+b-off}),
when the two sites of real-space coordinates $\vec{r}_1$ and $\vec{r}_2$ 
(and $\vec{r}_{1'}$ and $\vec{r}_{2'}$) are linked by a bond (and
occupied by two independent $+1/2$ spinons)
one must consider both initial configurations where the sites
of real-space coordinates $\vec{r}_{1'}$ and $\vec{r}_{2'}$
(and $\vec{r}_1$ and $\vec{r}_2$) are (i) linked by a bond
and (ii) occupied by two independent $+1/2$ spinons.
In turn, when the two sites of real-space coordinates 
$\vec{r}_1$ and $\vec{r}_2$ (and $\vec{r}_{1'}$ and $\vec{r}_{2'}$) 
are occupied by two independent $+1/2$ spinons 
(and linked by a bond) one must consider only initial 
configurations where the sites
of real-space coordinates $\vec{r}_{1'}$ and $\vec{r}_{2'}$
(and $\vec{r}_1$ and $\vec{r}_2$) are linked by a bond
(and occupied by two independent $+1/2$ spinons).

It then follows from analysis of the operator expression on
the right-hand side of Eq. (\ref{anti-com-b+b-on}) that
when in the initial configuration the two sites $\vec{r}_1$ and 
$\vec{r}_2$ are occupied by two independent $+1/2$ spinons
the operator term $\sum_{i=1,2}[1/2](1/2+s^z_{\vec{r}_i})^2$
transforms that configuration onto itself whereas the
remaining operator terms give zero. In turn, when in the initial 
configuration the two sites $\vec{r}_1$ and $\vec{r}_2$
are linked and correspond to an one-bond configuration
the operator terms
$\sum_{i=1,2}[1/2]\{(1/2+s^z_{\vec{r}_i})^2
-s^+_{\vec{r}_i}s^-_{\vec{r}_{\bar{i}}}(1/2+s^z_{\vec{r}_{\bar{i}}})\}$
transform that configuration onto itself whereas the
remaining operator terms give zero. 
On the other hand, when acting on the above initial configurations 
the operator on the right-hand side of Eq.
(\ref{anti-com-b+b-off}) gives always zero so that when acting
onto the subspace that the operators of Eqs. (\ref{g-s1+general})-(\ref{g-s-l})
refer to the anti-commutators provided in Eqs. (\ref{anti-com-b+b-on})
and (\ref{anti-com-b+b-off}) simplify and are given by,
\begin{equation}
\{b_{\vec{r}_1,\vec{r}_2}^{\dag},b_{\vec{r}_1,\vec{r}_2}\} = 1 \, ;
\hspace{0.50cm}
\{b_{\vec{r}_1,\vec{r}_2}^{\dag},b_{\vec{r}_{1'},\vec{r}_{2'}}\} = 0 \, ,
\label{anti-com-b+b-on-off}
\end{equation}
where $\vec{r}_1\neq \vec{r}_{1'}, \vec{r}_{2'}$ and
$\vec{r}_2\neq \vec{r}_{1'}, \vec{r}_{2'}$.

Next concerning the two general anti-commutators needed to evaluate the 
two-site one-bond operators on the right-hand side of Eq. (\ref{g-local-AP++}) 
we find the following expressions,
\begin{equation}
\{b_{\vec{r}_1,\vec{r}_2}^{\dag},b_{\vec{r}_1,\vec{r}_2}^{\dag}\} =
-2\left({1\over 2}+s^z_{\vec{r}_1}\right)\left({1\over 2}+s^z_{\vec{r}_2}\right)
s^-_{\vec{r}_1}s^-_{\vec{r}_2} -
\sum_{i=1,2}\left({1\over 2}+s^z_{\vec{r}_i}\right)s^-_{\vec{r}_1}s^-_{\vec{r}_2} \, ,
\label{anti-com-b+b+on}
\end{equation}
\begin{eqnarray}
\{b_{\vec{r}_1,\vec{r}_2}^{\dag},b_{\vec{r}_{1'},\vec{r}_{2'}}^{\dag}\} & = & (-1)^{d+d'}
\sum_{i=1,2}\{\left({1\over 2}+s^z_{\vec{r}_{i}}\right)
\left({1\over 2}+s^z_{\vec{r}_{i'}}\right)s^-_{\vec{r}_{\bar{i}}}s^-_{\vec{r}_{\bar{i}'}} 
\nonumber \\
& - & \left({1\over 2}+s^z_{\vec{r}_{i}}\right)
\left({1\over 2}+s^z_{\vec{r}_{\bar{i}'}}\right)s^-_{\vec{r}_{\bar{i}}}s^-_{\vec{r}_{i'}}\} \, ,
\label{anti-com-b+b+off}
\end{eqnarray}
where as above $\bar{1}=2$, $\bar{2}=1$, $\vec{r}_1\neq \vec{r}_{1'}, \vec{r}_{2'}$, and
$\vec{r}_2\neq \vec{r}_{1'}, \vec{r}_{2'}$. 

It follows from analysis of the operator on
the right-hand side of Eq. (\ref{anti-com-b+b+on}) that
according to the first and third rules 
when in the initial spin configuration the sites of
real-space coordinates $\vec{r}_1$ and 
$\vec{r}_2$ are both (i) occupied by independent $+1/2$ spinons
and (ii) linked by a bond, application of that operator onto
such a configuration gives zero. In turn the
anti-commutator of Eq. (\ref{anti-com-b+b+off}) is of the
form of those on the right-hand side of
Eq. (\ref{g-local-AP++}) so that in Eq. (\ref{variables}) one
has that $j=j'$ yet $\vec{r}_{d,l}^{\,g}\neq \vec{r}_{d',l'}^{\,g'}$
and then the forth rule applies. Since we find that
$b_{\vec{r}_1,\vec{r}_2}^{\dag}b_{\vec{r}_{1'},\vec{r}_{2'}}^{\dag}=
b_{\vec{r}_{1'},\vec{r}_{2'}}^{\dag}b_{\vec{r}_1,\vec{r}_2}^{\dag}$ 
and each of such operators is given by one half the
operator on the right-hand side of Eq. (\ref{anti-com-b+b+off}), concerning
the latter operator when the sites of real-space 
coordinates $\vec{r}_1$ and $\vec{r}_2$ (and $\vec{r}_{1'}$ and $\vec{r}_{2'}$)
are linked by a bond one must consider both initial configurations where the 
sites of real-space coordinates $\vec{r}_{1'}$ and $\vec{r}_{2'}$
(and $\vec{r}_1$ and $\vec{r}_2$) are (i) linked by a bond
and (ii) occupied by two independent $+1/2$ spinons.
In turn, when the two sites of real-space coordinates 
$\vec{r}_1$ and $\vec{r}_2$ (and $\vec{r}_{1'}$ and $\vec{r}_{2'}$) 
are occupied by two independent $+1/2$ spinons 
one must consider only initial configurations where the sites
of real-space coordinates $\vec{r}_{1'}$ and $\vec{r}_{2'}$
(and $\vec{r}_1$ and $\vec{r}_2$) are linked by a bond \cite{companion}. 
It follows then 
from analysis of the operator on the right-hand side of Eq. (\ref{anti-com-b+b+off}) 
that application of it onto any of such spin configurations gives zero.

A similar analysis for the two general anti-commutators needed to evaluate the 
two-site one-bond operators on the right-hand side of Eq. (\ref{g-local-AP--})
leads to, 
\begin{equation}
\{b_{\vec{r}_1,\vec{r}_2},b_{\vec{r}_1,\vec{r}_2}\} =
-2s^+_{\vec{r}_1}s^+_{\vec{r}_2}\left({1\over 2}+s^z_{\vec{r}_1}\right)\left({1\over 2}+s^z_{\vec{r}_2}\right) -
\sum_{i=1,2}s^+_{\vec{r}_1}s^+_{\vec{r}_2}\left({1\over 2}+s^z_{\vec{r}_i}\right) \, ,
\label{anti-com-b-b-on}
\end{equation}
\begin{eqnarray}
\{b_{\vec{r}_1,\vec{r}_2},b_{\vec{r}_{1'},\vec{r}_{2'}}\} & = & (-1)^{d+d'}
\sum_{i=1,2}\{s^+_{\vec{r}_{\bar{i}}}s^+_{\vec{r}_{\bar{i}'}}\left({1\over 2}+s^z_{\vec{r}_{i}}\right)
\left({1\over 2}+s^z_{\vec{r}_{i'}}\right) 
\nonumber \\
& - & s^+_{\vec{r}_{\bar{i}}}s^+_{\vec{r}_{i'}}\left({1\over 2}+s^z_{\vec{r}_{i}}\right)
\left({1\over 2}+s^z_{\vec{r}_{\bar{i}'}}\right)\} \, ,
\label{anti-com-b-b-off}
\end{eqnarray}
where as above $\bar{1}=2$, $\bar{2}=1$, $\vec{r}_1\neq \vec{r}_{1'}, \vec{r}_{2'}$, and
$\vec{r}_2\neq \vec{r}_{1'}, \vec{r}_{2'}$. 

Again analysis of the operator on
the right-hand side of Eq. (\ref{anti-com-b-b-on}) reveals that
according to the first and third rules 
when in the initial spin configuration the sites of
real-space coordinates $\vec{r}_1$ and 
$\vec{r}_2$ are both (i) occupied by independent $+1/2$ spinons
and (ii) linked by a bond, application of that operator onto
such a configuration gives zero. On the other hand, the
anti-commutator of Eq. (\ref{anti-com-b-b-off}) is of the
form of those on the right-hand side of
Eq. (\ref{g-local-AP++}) so that in Eq. (\ref{variables}) one
has that $j=j'$ yet $\vec{r}_{d,l}^{\,g}\neq \vec{r}_{d',l'}^{\,g'}$
and then the forth rule applies. Since we find that
$b_{\vec{r}_1,\vec{r}_2}b_{\vec{r}_{1'},\vec{r}_{2'}}=
b_{\vec{r}_{1'},\vec{r}_{2'}}b_{\vec{r}_1,\vec{r}_2}$ 
and each of such operators is given by one half the
operator on the right-hand side of Eq. (\ref{anti-com-b+b+off}), concerning
the latter operator when the sites of real-space 
coordinates $\vec{r}_1$ and $\vec{r}_2$ (and $\vec{r}_{1'}$ and $\vec{r}_{2'}$)
are occupied by two independent $+1/2$ spinons one must 
consider both initial configurations where the 
sites of real-space coordinates $\vec{r}_{1'}$ and $\vec{r}_{2'}$
(and $\vec{r}_1$ and $\vec{r}_2$) are (i) linked by a bond
and (ii) occupied by two independent $+1/2$ spinons.
In turn, when the two sites of real-space coordinates 
$\vec{r}_1$ and $\vec{r}_2$ (and $\vec{r}_{1'}$ and $\vec{r}_{2'}$) 
are linked by a bond one must consider only initial configurations where the sites
of real-space coordinates $\vec{r}_{1'}$ and $\vec{r}_{2'}$
(and $\vec{r}_1$ and $\vec{r}_2$) are occupied by two 
independent $+1/2$ spinons \cite{companion}. 
Analysis of the operator on the right-hand side of Eq. (\ref{anti-com-b+b+off}) 
then reveals that application of it onto any of such spin configurations gives zero.

It then follows from the above results that when acting
onto the subspace that the operators of Eqs. (\ref{g-s1+general})-(\ref{g-s-l})
refer to the anti-commutators provided in Eqs. (\ref{anti-com-b+b+on})
and (\ref{anti-com-b+b+off}) and Eqs. (\ref{anti-com-b-b-on})
and (\ref{anti-com-b-b-off}) simplify and read,
\begin{equation}
\{b_{\vec{r}_1,\vec{r}_2}^{\dag},b_{\vec{r}_1,\vec{r}_2}^{\dag}\} =
\{b_{\vec{r}_1,\vec{r}_2}^{\dag},b_{\vec{r}_{1'},\vec{r}_{2'}}^{\dag}\} = 0 
\, ; \hspace{0.5cm}
\{b_{\vec{r}_1,\vec{r}_2},b_{\vec{r}_1,\vec{r}_2}\} =
\{b_{\vec{r}_1,\vec{r}_2},b_{\vec{r}_{1'},\vec{r}_{2'}}\} = 0 \, ,
\label{anti-com-b+-b+-on-off}
\end{equation}
where $\vec{r}_1\neq \vec{r}_{1'}, \vec{r}_{2'}$ and
$\vec{r}_2\neq \vec{r}_{1'}, \vec{r}_{2'}$.

The use in Eq. (\ref{g-local-AP+-}) of the anti-commutators of Eq. (\ref{anti-com-b+b-on-off}) with 
the two-site one-bond operators related to those of Eq. (\ref{g-s-l}) by 
the expressions provided in Eqs. (\ref{g-s-l-simple}) and (\ref{variables}) leads to,
\begin{equation}
\{g^{\dag}_{{\vec{r}}_{j},s1},g_{{\vec{r}}_{j},s1}\} =
\sum_{d,l,g} \vert h_g\vert^2 = 2D\sum_g \vert h_g\vert^2 =1 \, ,
\label{g-local-AP+-final}
\end{equation}
which is the first relation of Eq. (\ref{g-local}). To perform the
summation of Eq. (\ref{g-local-AP+-final}) the sum-rule (\ref{g-s1+sum-rule}) 
was used. Furthermore, the use of the anti-commutators of Eq. (\ref{anti-com-b+-b+-on-off})
in Eqs. (\ref{g-local-AP++}) and (\ref{g-local-AP--}) leads to the remaining
relations of Eq. (\ref{g-local}).

The evaluation of the commutators of Eq. (\ref{g-non-local}) by the use of the 
expressions given in Eqs. (\ref{g-non-local-AP+-})-(\ref{g-non-local-AP--})
is much simpler. First it is simplified by the property that two-site one-link bonds belonging
to $s1$ bond-particle operators with different real-space coordinates
are always different. Second the evaluation of such commutators also relies 
on straightforward manipulations based on Eqs. 
(\ref{albegra-s-p-m})-(\ref{albegra-s-sz-com}), which lead directly to,
\begin{equation}
[b_{\vec{r}_1,\vec{r}_2}^{\dag},b_{\vec{r}_{1'},\vec{r}_{2'}}] =
[b_{\vec{r}_1,\vec{r}_2}^{\dag},b_{\vec{r}_{1'},\vec{r}_{2'}}^{\dag}] =
[b_{\vec{r}_1,\vec{r}_2},b_{\vec{r}_{1'},\vec{r}_{2'}}] = 0 \, ,
\label{com-all}
\end{equation}
for $\vec{r}_1\neq \vec{r}_{1'}, \vec{r}_{2'}$ and
$\vec{r}_2\neq \vec{r}_{1'}, \vec{r}_{2'}$.

Finally, the use in the expressions of the $s1$ bond-particle operators of
Eq. (\ref{g-non-local}) of the commutators of Eq. (\ref{com-all}) with 
the two-site one-bond operators related to those of Eq. (\ref{g-s-l}) by 
the expressions provided in Eqs. (\ref{g-s-l-simple}) and (\ref{variables}) 
in Eqs.  (\ref{g-non-local-AP+-})-(\ref{g-non-local-AP--}) leads readily
to the commutation relations provided in Eq. (\ref{g-non-local}).


\end{document}